%% file: main_paper_CC_2012.tex
\documentclass[times]{cpeauth}

\usepackage{moreverb}

\usepackage[colorlinks,bookmarksopen,bookmarksnumbered,citecolor=red,urlcolor=red]{hyperref}

\newcommand\BibTeX{{\rmfamily B\kern-.05em \textsc{i\kern-.025em b}\kern-.08em
T\kern-.1667em\lower.7ex\hbox{E}\kern-.125emX}}

\def\volumeyear{2012}

\begin{document}

\runningheads{Arnaldo, Cuesta-Infante, Colmenar, Risco-Mart\'in, Ayala}{Boosting the 3D Thermal-Aware Floorplaning Problem through a M-W PMOEA}

\title{Boosting the 3D Thermal-Aware Floorplanning Problem through a Master-Worker Parallel MOEA}


\author{Ignacio Arnaldo\affil{1},
Alfredo Cuesta-Infante\affil{2},
J. Manuel Colmenar\affil{2}\corrauth,
Jos\'e L. Risco-Mart\'in\affil{1},
Jos\'e L. Ayala\affil{1}}

\address{\affilnum{1}DACYA, Complutense University of Madrid\break
\affilnum{2}CES Felipe II, Aranjuez, Complutense University of Madrid}

\corraddr{C/. Capit\'an, 39, 28300, Aranjuez (Madrid), Spain. E-mail: jmcolmenar@ajz.ucm.es}


\cgs{Ignacio Arnaldo is supported by Spanish Government Avanza Competitividad I+D+I: TSI-020100-2010-962 project. The work has also been supported by Spanish Government grants TIN 2008-00508 and MEC CONSOLIDER CSD00C-07-20811}

\begin{abstract}
\vspace{0.4cm}
The increasing transistor scale integration poses, among others, the thermal-aware floorplanning problem; consisting of how to place the hardware components in order to reduce overheating by dissipation.
Due to the huge amount of feasible floorplans, most of the solutions found in the literature include an evolutionary algorithm for, either partially or completely, carrying out the task of floorplanning. 
Evolutionary algorithms usually have a bottleneck in the fitness evaluation. 
In the problem of thermal-aware floorplanning, the layout evaluation by the thermal model takes 99.5\% of the computational time for the best floorplanning algorithm proposed so far.
The contribution of this paper is to present a parallelization of this evaluation phase in a master$-$worker model to achieve a dramatic speed-up of the thermal-aware floorplanning process.
Exhaustive experimentation was done over three dimensional integrated circuits, with 48 and 128 cores, outperforming previous published works.
\end{abstract}




\maketitle

%

\input{01_intro}

\input{02_core}
 
\input{03_results}

 
\input{04_conclusions_future}
\bibliographystyle{wileyj}
\bibliography{main_paper_CC_2012}  

\end{document}

%% file: 01_intro.tex
\section{Introduction}
\vspace{0.4cm}
Consumers continuously demand faster applications, smaller devices and recently also ubiquitous computing. 
So far, developments in materials and technology have allowed processor manufacturers to provide chips that attained the expected serial performance.
As we approach to the limits of miniaturization, these demands become harder to accomplish.
In order to remain competitive, industry has moved to parallel architectures such as integrating more cores in a die, data-parallel execution units, additional register sets for hardware threads, bigger caches and more independent memory controllers to increase memory bandwidth.
For instance, multi-core general purpose computers are being shipped for years and 
data-centers implement heterogeneous many-core systems.
Multi-processor systems-on-chip (MPSoCs) are nowadays also considered as many-core systems.
Up to now, the top core integration silicon CPU chip is proposed by Intel Labs with an experimental Single-chip Cloud Computer (SCC), a research microprocessor containing 48 Intel Architecture cores \cite{intelscc}.
Also, novel 3D multi-processor chips have been recently presented \cite{Loh2010} as the alternative to provide the required area of integration and to reduce the communication delay among the large number of cores. 

While the fabrication techniques have driven the integration of an increased number of transistors to provide the required throughput, these improvements have posed major problems regarding the operating temperature, directly related to the power density \cite{Borkar1999}. 
As temperature increases, the carrier mobility degrades, the leakage power consumption increases, gradient temperatures appear on the surface creating electro-migrations and the lifetime of the chip decreases exponentially, all in all reducing dramatically the reliability of the system \cite{Srinivasan2004}. 
In addition, the specific placement of the functional units on the chip surface (floorplan) also affects to the temperature distribution because of the diffusive nature of heat \cite{Sanka2005}. 
Besides, in the 3D configuration, the power density increases with the number of layers. 
This effect is even more negative due to the problematic cooling of inner layers of the 3D stack.
The impact of power density in the microprocessor is augmented due to the dielectric insulation layers inserted between active layers. The reason is that the thermal conductivity of the formers is very low compared to silicon and metal. 

Increasing the chip area to reduce the power density has two shortcomings: it is costly and requires to solve all the geometric constraints. 
Static external coolers reduce the temperature of the chip surface by a constant factor but do not reduce the temperature gradient across the chip. 
Instead, thermal-aware floorplanning algorithms attempt to place functional units in order to achieve a satisfactory temperature distribution; thus they tackle with both heat dissipation and component placement at a time.

Floorplanning proposals are frequently formulated as combinatorial optimization problems that can be smoothly fit to genetic algorithms (GA).

Broadly speaking, a GA performs a heuristic search throughout the solution space inspired on Darwin's principle of Natural Selection:
The basic features are:\\
(i) each iteration tests a small number of solutions (known as \textit{population}) compared to the cardinality of the solution space,\\
(ii) each solution (referred to as \textit{individual}) is represented in a way suitable both for evaluation and for producing the subset of the next generation,\\
(iii) the next generation is obtained applying genetic operators such as crossover,  mutation and selection, and\\
(iv) there is a fitness function that evaluates the individuals.

Early floorplanning solutions tackled with GA proposed representations such as Polish notation \cite{Berntsson2004}, combined bucket array \cite{Cong2004} and O-trees \cite{Tang2007} that are not satisfactory in the thermal-aware problem because they were engineered to minimize area. On the contrary, the hottest elements should be spread and placed as far as possible for reducing the heating produced by closer hot units. 
Thus, in 2D, works like \cite{Hung2005} decreased the peak temperature using genetic algorithms, and \cite{Han2007} using simulated annealing; whereas on 3D stacked systems linear programming and simulated annealing combinations may be found \cite{Healy2007}.

These works solve a single-objective optimization that takes into account only the impact that temperature has on reliability. Therefore, they cannot provide a thermally optimal solution with a minimum impact on the area of the chip and the delay due to wiring. 
A more comprehensive approach in 3D systems is presented in \cite{Cuesta2011b} and \cite{risco2011}. Their proposal consists of a Multi-Objective Evolutionary Algorithm (MOEA) that tackles with the thermal-aware problem (optimal placement of blocks) and also with the performance of the system (minimum wire length delay) satisfying the topological constraints. On the other hand, they show a critical bottleneck in the evaluation phase due to the complexity of the computation, and can lead to very long execution times when complex 3D many-cores architectures are considered. 


\begin{table}
\caption{Execution time of the different methods involved in the algorithm proposed in \cite{Cuesta2011b}.}
\label{computationalLoads}
\centering
\tabsize
\begin{tabular}{cc}
\toprule
Method/Operator & Time (ms)\\
\midrule
Evaluation      & 31264762\\
Selection       & 257 \\
Reduction       & 251 \\
Mutation        & 57.6 \\
Crossover       & 1 \\
\bottomrule
\end{tabular}
\end{table}


Our contribution in this paper is the parallelization of the thermal-aware floorplanner proposed in \cite{Cuesta2011b} and \cite{risco2011}, with the aim of reducing the optimization execution time. 

Evolutionary algorithms (EA) are intrinsically parallel but it is in the fitness evaluation where more speed-up can be gained. 
Table \ref{computationalLoads} shows the complete execution time of the sequential version of the algorithm proposed in \cite{Cuesta2011b} until a solution is reached. The evaluation and reduction phases correspond to methods devoted to compute the fitness and ranking of the individuals, the rest implement the genetic operators of selection, crossover and mutation. 
In this case, fitness is computed using a thermal model that takes $83.1\%$ of the execution time. Due to the simple representation chosen for the candidate solutions, a decodification phase is required before the fitness computation. Adding the decodification phase time and the feasibility verification time, the  evaluation of individuals takes $99.5\%$ of the total execution time. It is then clear that this task is by far the most time consuming which justifies the necessity and effort of parallelization.

An EA is usually parallelized at two different levels: definition of population or fitness evaluation \cite{Cantu98}. 
At the former, the population is split in a number of non-overlapping subpopulations that evolve independently but with a probability of interaction. The two most popular models are Islands and Grid models \cite{Cantu98}. 
In Islands, some individuals are allowed to migrate with a given frequency. There is a rich variety of Islands topologies, being the most frequent rings, $n$-dimensional meshes and stars. When migrating, the worst $k$ individuals in the destination are replaced by the new-comers, which are the best $k$ individuals in their original island.
In grids, each individual is placed in a cell of a one- or two-dimensional grid.
Genetic operations take place in a small neighborhood of a given individual and their implementation is straightforward on clusters.

The fitness evaluation parallelization is a much simpler and intuitive approach. All the genetic operations are performed sequentially over the whole population but, once a new generation is obtained, the individuals are evaluated in parallel.

Regarding parallel MOEAs; they were early analyzed in \cite{VanVeldhuizen2003}. 
Shortly afterwards, the master-slave paradigm was employed in \cite{LopezJaimes2005}.

Since our baseline sequential algorithm presents a high computational load in the evaluation of individuals, this paper proposes a master-worker algorithm to parallelize that phase.
Our approach was tested with a set of master-worker configurations, ranging from 1 to 9 workers, as well as the sequential algorithm, in two experimental multi-core platforms with 48 and 128 3D stacked core processors each. Speedup review, validation of the proposal, study of convergence and thermal analysis are presented in this paper. 
Results suggest that a new representation could improve future algorithms.


The rest of the paper is organized as follows.
The parallel Multi-Objective Evolutionary Algorithm proposal is presented in Section 2.
Experimental results are shown and discussed in Section 3.
Finally, conclusions and future work are detailed in Section 4.

%% file: 02_core.tex
\section{Parallel Multi-Objective Evolutionary Algorithm}
\vspace{0.4cm}
In this section we present a parallel thermal-aware floorplanner capable of optimizing many-core heterogeneous platforms under a master-worker schema.

\subsection{Details of the MOEA}
\vspace{0.4cm}

In \cite{Cuesta2011b}, a Multi-Objective Evolutionary Algorithm (MOEA) is proposed for the floorplanning of 3D stacked multi-processor single-chips.
This kind of chips consists of a number of layers of fixed area where the functional units (processors, memories and interconnection blocks) must be placed. Our approach considers this scenario performing a thermal-aware placement of the different components while the wiring delay is also minimized. Moreover, it overcomes \cite{Cuesta2011b} with a parallel implementation of the floorplanner that avoids the constraints imposed to the optimization time.

\begin{figure}[t]
\centering
\includegraphics[width=0.65\textwidth]{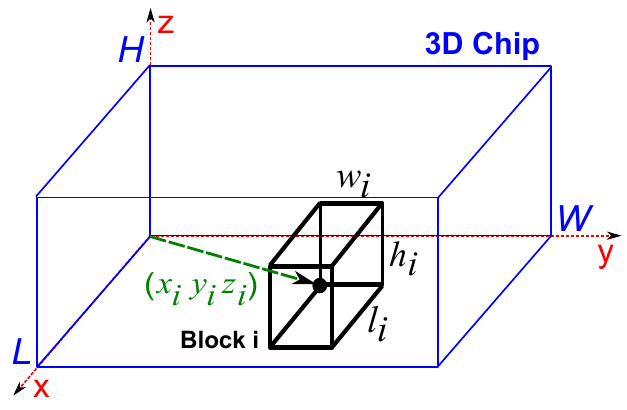}
\caption{Block representation.}
\label{fig:block}
\end{figure}

\vspace{0.4cm}
\textbf{Block placement problem.~}
All the components that model the functional units of the many-core system must be placed in the 3D stack, which imposes the physical boundaries of maximum length $L$, width $W$ and height $H$. 
Each component is represented by a block $B_i$ with length $l_i$, width $w_i$, height $h_i$ and it is denoted by its left-bottom-back corner, with coordinates $(x_i,y_i,z_i)$, taken the left-bottom-back corner of the chip as origin of coordinates; where
$0 \le x_i \le L-l_i$, $0 \le y_i \le W-w_i$, $0 \le z_i \le H-h_i$.
These blocks cannot overlap. 
A schematic representation is shown in Figure \ref{fig:block}. 

Blocks are placed sequentially. Since each component incorporates its coordinates, this method leads to a floorplan whose components are not necessarilly adjacent; unlike the state of the art works \cite{Berntsson2004}, \cite{Cong2004} or \cite{Tang2007}. 
This represents a great advantage because cores, which are the most likely to increase the temperature, can be placed explicitly separated, thus reducing their impact in the overall temperature of the 3D chip. 
In order to select the best placement coordinate $r_i=(x_i,y_i,z_i)$ for block $B_i$, given those already placed $B_j, j<i$, a dominance relationship is established. Therefore, a set of objective functions that evaluate the fitness, as well as a suitable representation and appropriate genetic operators, must be derived for the MOEA approach. The solution is obtained using a Non-dominated Sorting Genetic Algorithm (NSGA-II) implementation \cite{Deb2002}.

\vspace{0.4cm}
\textbf{Fitness.~}
There are three objective functions. 
The first objective $J_1$ is the number of topological constraints violated by $B_i$ with respect to the already placed $B_j$.
The second objective is the wire length, approximated by the Manhattan distance between blocks with coordinates $r_i$ and $r_j$,
$J_2=|x_i-x_j|+|y_i-y_j|+|z_i-z_j|.$
Finally the thermal impact is measured through the power consumed by the unitary cells of the chip. A thermal model that considers the power density of such cells and their neighbors is used as an approximation to the steady state of the more accurate thermal model that includes non-linear and differential equations. We evaluate the thermal response of a given individual with the following model:
\begin{equation}
J_ 3 = \sum_{i<j \in 1..n}(dp_i*dp_j)/(d_{ij})
\end{equation}
where $dp$ is the density power of the block considered and $d_{ij}$ is the euclidean distance between blocks. This model has been shown to be accurate enough and close to the non-linear simulation \cite{Paci2007}.

\vspace{0.4cm}
\textbf{Representation.~} 
Each individual is a possible configuration of the system, and each configuration is represented through a chromosome that stores the order in which blocks are being placed. 
The chromosome is an array whose components contain an identifier of the functional unit that is going to be placed. 
For instance, if 3 cores $(C_1,C_2,C_3)$ and 3 memories $(L_1,L_2,L_3)$ are to be placed, the search space will have cardinality $6!$. A possible chromosome would be $$[C_1,L_2,C_3,L_3,C_2,L_1]$$
meaning that $C_1$ will be fist placed, then $L_2$ and so on.

This decodification of the chromosome requires considering the size and boundaries of the previously placed blocks. Therefore, the more the number of components, the more the decodification execution time. As we will describe in the future work section of the paper, a different representation encoding the location of the components could help reducing the evaluation phase.

\vspace{0.4cm}
\textbf{Operators.~}
\begin{figure}[t]
	\centering
	\includegraphics[width=0.65\textwidth]{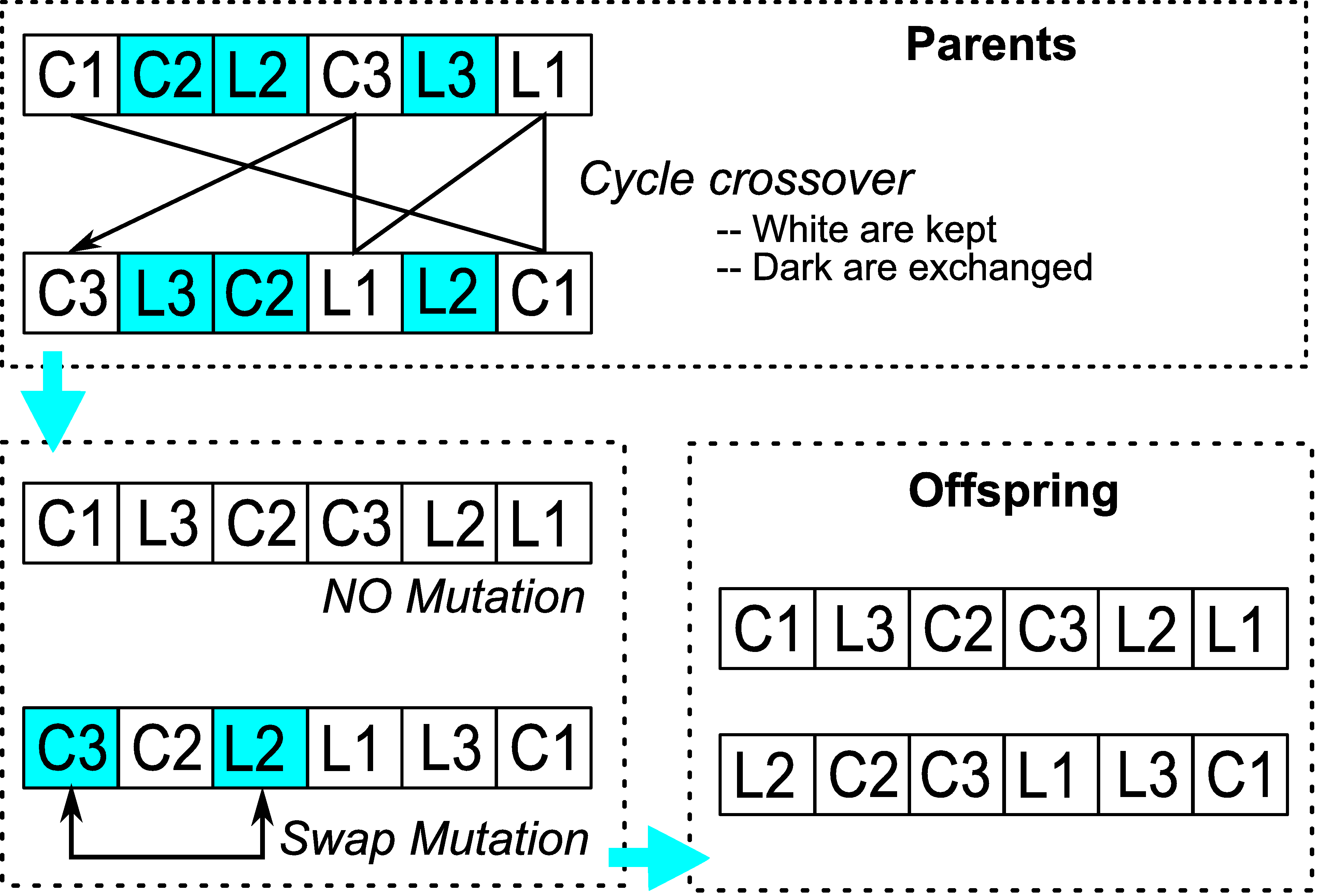}
	\caption{Cycle crossover and Swap mutation for permutations of six elements.}
	\label{fig:geneticOP1}
\end{figure}
Selection was carried out by tournament, taking two random individuals and selecting the best among them. 
Individuals are mated  in order to produce offspring.
Crossover must take into account that all the components must appear once and only once in the chromosome. The so called \emph{cycle crossover} assures that the resulting chromosomes are just permutations of the parents.
Mutation consists of swapping the content of two positions inside the chromosome or in the rotation of a randomly chosen component.
Both cycle crossover and swap mutation are depicted in Figure \ref{fig:geneticOP1}.

\subsection{Details of Parallelization}
\vspace{0.4cm}
In this paper, we propose a parallel implementation of the MOEA described in the previous section using the master-worker model. As we have previously shown, the evaluation phase of the algorithm takes over 99\% of the execution time. 
This is due to both the fact that the thermal response of all the individuals of the population has to be evaluated in every generation of the process, and the decodification of each individual, previous to its evaluation. 

\begin{figure}[t]
\centering
\includegraphics[width=0.5\textwidth]{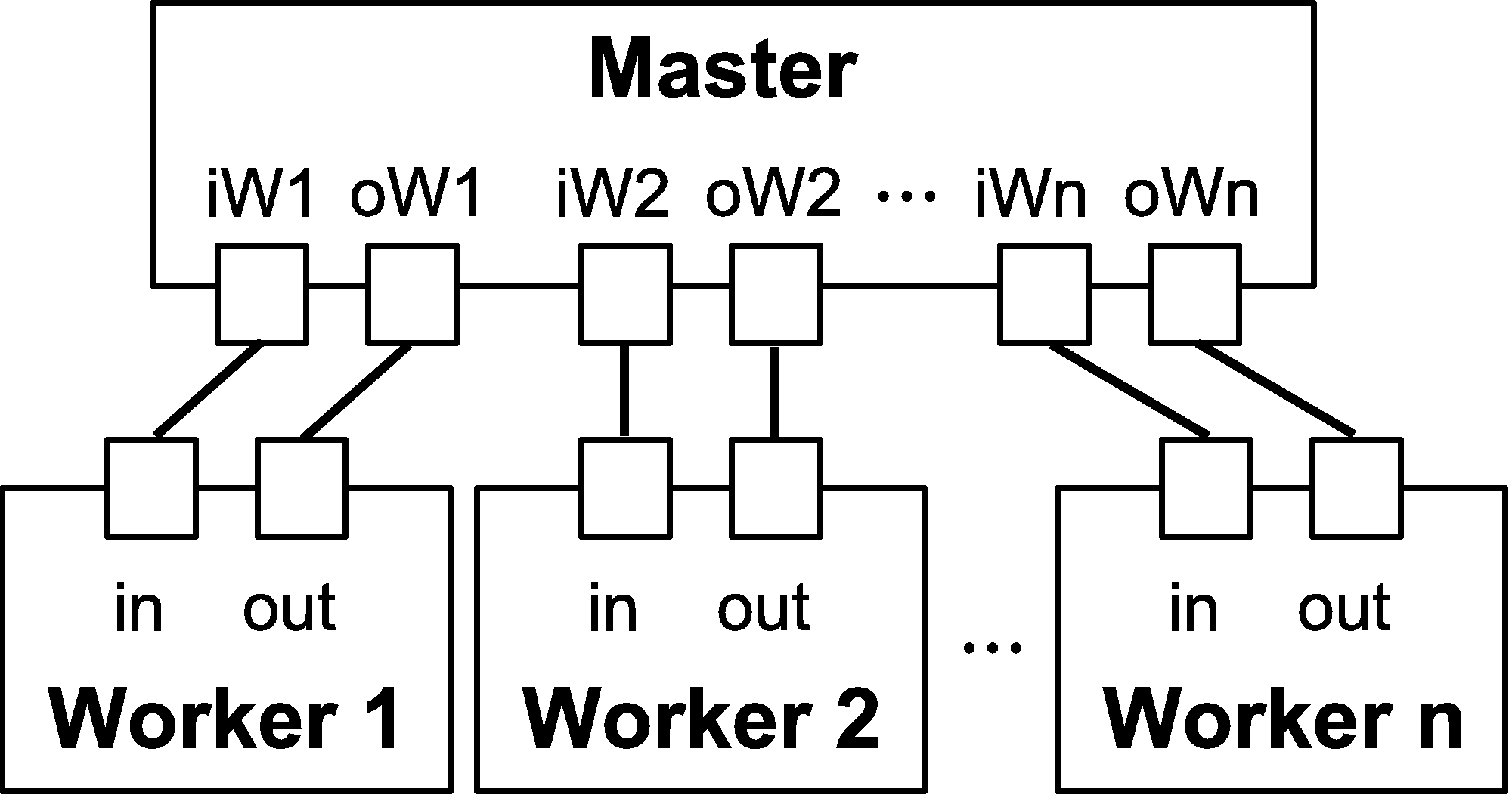}
\caption{Master-worker configuration.}
\label{fig:MasterWorker}
\end{figure}

The master-worker model satisfies our needs because, even though the fitness is based on a simplified thermal model, the computational cost of this evaluation increases quadratically with the number of components. Therefore, it is interesting to exploit the fact that evolutionary algorithms are intrinsically parallel and carry out the evaluation of the population in a concurrent manner. Figure \ref{fig:MasterWorker} depicts the approach used in this work. 
The master distributes the population among $n$ workers, splitting the computational load in $n$ ways so it does not carry out any evaluation.
Once workers have finished their task, they send the outcome together with the received population subset to the master. Although the algorithm stops and waits for all workers to finish, it is clearly much faster than the sequential execution as long as each subset was large enough for compensating communication times. 

We propose a multi-threaded implementation where only the master executes the main thread of the algorithm. 
Since only workers execute the evaluation of different subsets of the population, it is expected to obtain a speed-up similar to the number of cores in the processor that executes the algorithm. 

%% file: 03_results.tex
\section{Experimental results}
\vspace{0.4cm}
The experimental work will analyze the speedup obtained with the parallel version of the MOEA while making clear that the quality of the solutions remains the same. 
We will also analyze the thermal optimization achieved by the floorplanner. 

In order to evaluate our floorplanner, we study two heterogeneous 3D architectures where every stacked layer is based on the Niagara platform. They differ, from each other, in the number of cores. The first architecture is composed of 48 processor cores: 32 SPARC and 12 Power6 cores. Adding 72 memories and 6 crossbar for inter-processor communication they sum up a total of 126 components. In the second architecture, 128 cores are included: 96 SPARC plus 32 Power6. In addition, 192 memories and 16 crossbars are considered, therefore 336 components need to be placed in this scenario. The floorplanner will place the processors, the local memories and the crossbars in 4 and 9 layers respectively. Both architectures represent the current and the nearly future state-of-the-art in 3D many-core integration.

\subsection{Speedup Analysis}
\vspace{0.4cm}
In the first analysis, we studied the speedup obtained with the parallel version of our floorplanner. We aim to find the optimal number of workers leading to the maximum speedup. To this end, we perform a parametric sweep of the number of workers, from 1 to 9, both in the 48 and 128 core scenarios. In order to obtain the execution time of these optimizations, we run five times each one of the worker configurations, obtaining the average execution time and speedup for both scenarios. The experiments were carried out in a dedicated Intel Core-i5 machine, a 4-core processor, running at 2.80GHz. 

We set a fixed population size of 100 individuals and 250 generations evolution as the MOEA parameters for the 48 core scenario optimization. Table \ref{execs48} shows the average execution times and corresponding speedups for these runs, with a number of workers ranging from 1 to 9. 

\begin{table}
\caption{Average execution times and speedups obtained in the 48 cores scenario.}
\centering
\tabsize
\begin{tabular}{cccccccccc}
\toprule
\# workers & 1 & 2 & 3 & 4 & 5 & 6 & 7 & 8 & 9\\
\midrule
time (s) & 24171 & 12398 & 8904 & 6918 & 6380 & 6421 & 6665 & 6502 & 6386\\
speedup  & 1 & 1.95  & 2.72 & 3.49 & 3.79 & 3.76 & 3.63 & 3.72 & 3.79\\
\bottomrule
\end{tabular}
\label{execs48}
\end{table}

\begin{figure}
\centering
       \includegraphics[width=0.8\textwidth]{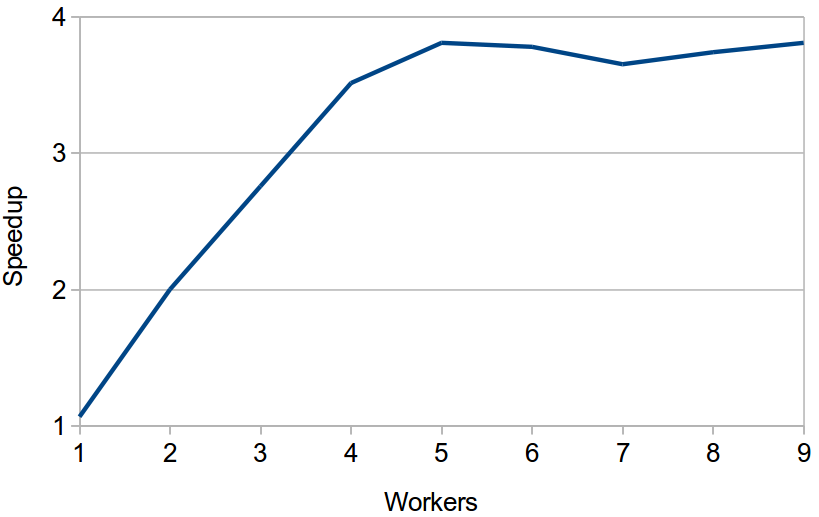}
\caption{Average speedup values obtained in the 48 cores scenario.}
\label{speedups48}
\end{figure}

Figure \ref{speedups48} shows the obtained speedups in the 48 core scenario. It shows see that the speedup increases almost linearly until the number of workers reaches the number of cores of the processor we used for these optimizations (4-core processor).

In our master-worker scheme, the execution time of each worker depends on the particular evaluation time of the set of individuals that was assigned to the worker. Then, if the worker receives a set of individuals that need more time to be evaluated, the worker will slow down. On the contrary, if the individuals need less time to be evaluated, the worker will speed up.

Therefore, in configurations from 2 to 4 workers, the optimization follows this behavior: the population is divided into as many sets as number of workers, then the set of individuals is sent to a different worker and the evaluation begins. Once workers finish their evaluation task, they wait until the slowest worker ends up, because the master synchronizes all the workers until the next generation. As a result, the slowest worker establishes the execution time of the evaluation of each generation, and the processor cores that run faster workers will be idle waiting for synchronization.

Each worker runs in a different thread, and the load assigned to one thread cannot be divided into different processor cores. As a consequence, if the number of workers is higher than 4, the operating system scheduler will distribute the execution of the worker threads among the 4 cores. Then, the cores will swap between the threads, therefore advancing on the execution of each one. As can be seen in Figure \ref{speedups48}, results for 5 workers and above present an asymptotic trend on speedup because the usage of resources is maximized. The particular case of the 5 workers configuration obtains the maximum speedup because maximizes the resources occupation with the lowest number of threads.

These results confirm that the parallelization of the evaluation phase in the master-worker scheme contributes to the best speedup gains. In addition, there is no remarkable penalty due to the parallelization, because the speedup values above 4 cores tend to be similar.



\begin{figure}
\centering
       \includegraphics[width=0.8\textwidth]{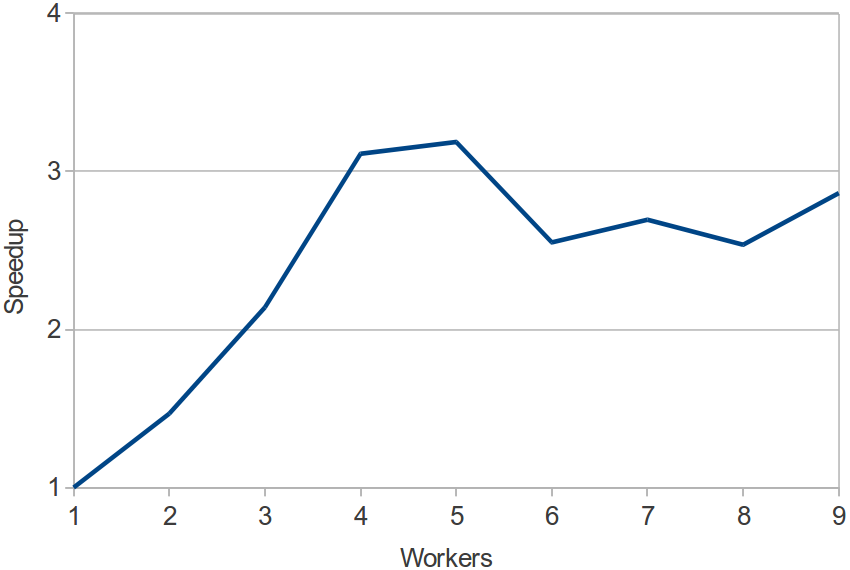}
\caption{Average speedup values obtained in the 128 cores scenario.}
\label{speedups128}
\end{figure}

In order to strengthen this hypothesis, the same tests were run for the 128 cores scenario, where the evaluation time for each individual is much  longer. Here, the number of generations of the MOEA has to be scaled up because the number of components to be placed has been increased. Therefore we consider a number of generations equal to the total number of components, which is 336: 128 cores, 192 memories and 16 crossbars. The population size remains 100 individuals. Table \ref{execs128} shows the average execution time and speedup for this scenario from 1 to 9 workers configuration. Figure \ref{speedups128} displays the speedup trend for these data.


\begin{table}
\caption{Average execution times and speedups obtained in the 128 cores scenario.}
\centering
\tabsize
\begin{tabular}{cccccccccc}
\toprule
\# workers & 1 & 2 & 3 & 4 & 5 & 6 & 7 & 8 & 9\\
\midrule
time (s) & 634070 & 432160 & 264318 & 203749 & 198977 & 248527 & 235313 & 250021 & 221485\\
speedup  & 1 & 1.47  & 2.14 & 3.11 & 3.19 & 2.55 & 2.69 & 2.54 & 2.86\\
\bottomrule
\end{tabular}
\label{execs128}
\end{table}



As was shown, the 128 cores scenario presents the same behavior as the 48 cores one. The resources of the CPU are maximized from the 5 workers configuration on, and higher numbers of workers obtain similar speedup values. However, the performance improvement is lower than in the 48 cores configuration. This behavior occurs because the individual evaluation time is much higher in this 128 cores scenario, and the execution time of the threads does not differ so much. In the 48 cores case, the processor slots available due to the different execution time between threads allow the evaluation of more individuals than in the 128 cores configuration. As a consequence, the workers queue, waiting for processor cores, advance more in their execution, obtaining higher speedups. On the contrary, the workers of the 128 cores scenario are not able to exploit the processor free slots to evaluate as many individuals as in the 48 cores case, therefore obtaining lower speedup values.

\subsection{Validation of solutions}
\vspace{0.4cm}

\begin{figure}[!h]
\centering
       \includegraphics[width=0.7\textwidth]{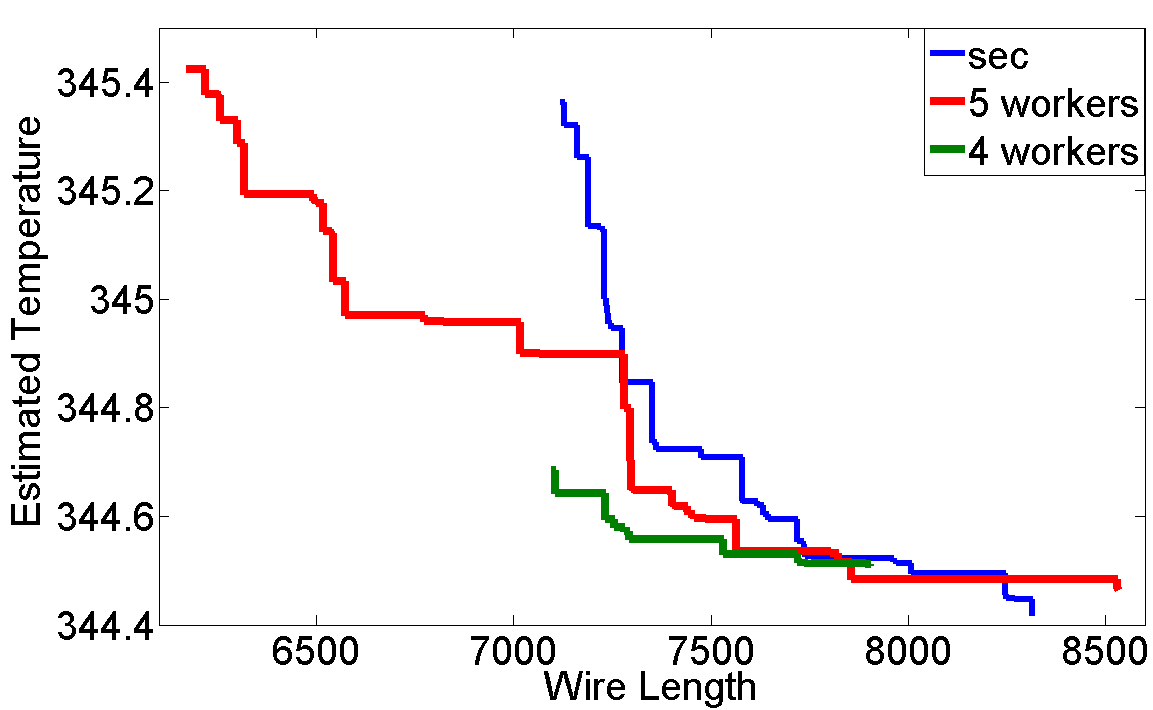}
\caption{Non-dominated fronts of solutions returned by the floorplanner in the 48 cores scenario.}
\label{fronts48}
\end{figure}

Conceptually it is clear that the parallelization of the fitness evaluation should lead to the same results than the sequential version of the algorithm. Nevertheless, we have included in this section a brief consideration regarding the validation of the parallelization that assures such a baseline. 
Thus, in order to show that the quality of the solutions proposed by the parallel version of the floorplanner remains the same than in the sequential version, we compare the front of non-dominated solutions obtained with the sequential version of the algorithm with the front obtained with the parallel version using 4 and 5 workers in the 48 cores scenario. Figure \ref{fronts48} shows the fronts of non-dominated solutions returned by the floorplanner in these cases.

\begin{figure}[!h]
\centering
       \includegraphics[width=0.7\textwidth]{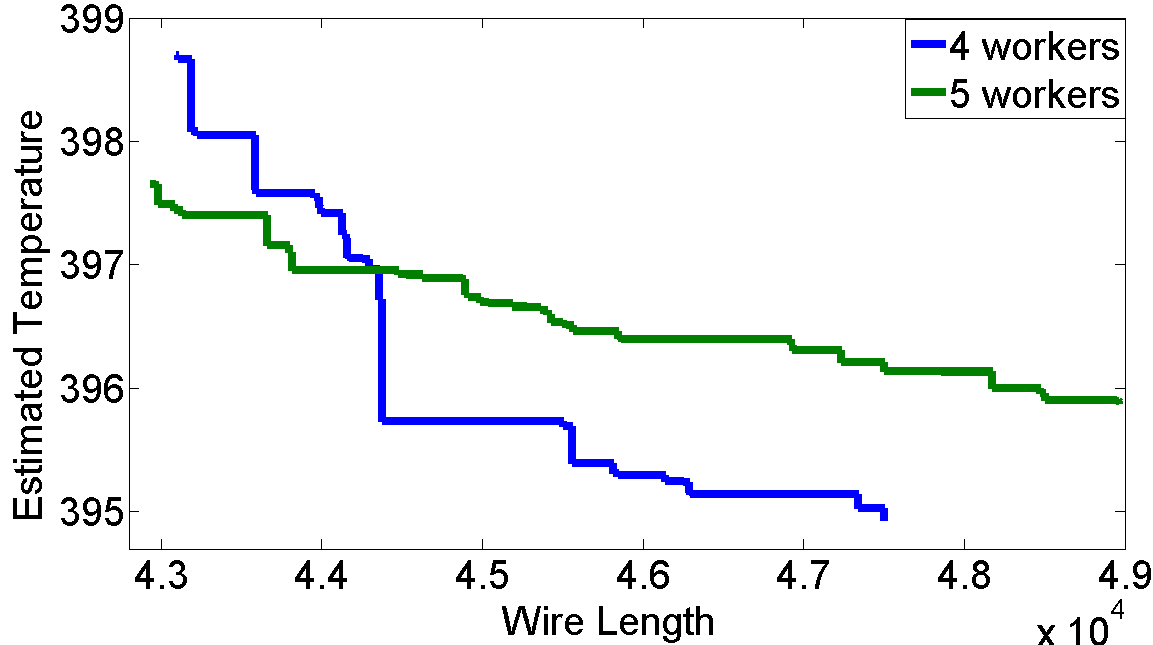}
\caption{Non-dominated fronts of solutions returned by the floorplanner in the 128 cores scenario.}
\label{fronts128}
\end{figure}

In Figure \ref{fronts128}, we compare the front of non-dominated solutions proposed by the floorplanner working with 4 and 5 nodes for the 128 cores platform.

For every run of the algorithm, the returned non-dominated front is different. In fact each execution explores a different region of the solution space. We can see that the fronts cross each other in at least one point. Therefore, none of the returned fronts dominates the others.


Since EA are intrinsically heuristic, two executions will not produce exactly the same results. 
Hence, in order to prove that our proposal is valid it is necessary to define a 
measure 
that analyzes the outputs (solution sets) both from sequential and parallel executions.
Such a measure is usually referred to as \emph{Indicator} $I$.  
In this work, the \emph{Hypervolume} indicator, proposed by Zitzler and Thiele \cite{Zitzler1999}, has been used.
The hypervolume $I(A)$ measures the total amount of the objective space that has been `covered' by the solution set $A$; returning the hypervolume of that portion of the objective space that is weakly dominated by $A$. 
To this end, the objective space must be bounded. 
Otherwise a reference point that must be at least weakly dominated by all solutions in $A$ is used.
Finally higher values of $I$ correspond to higher quality of the measured set.

\begin{figure}[t]
\centering
\includegraphics[width=.9\textwidth]{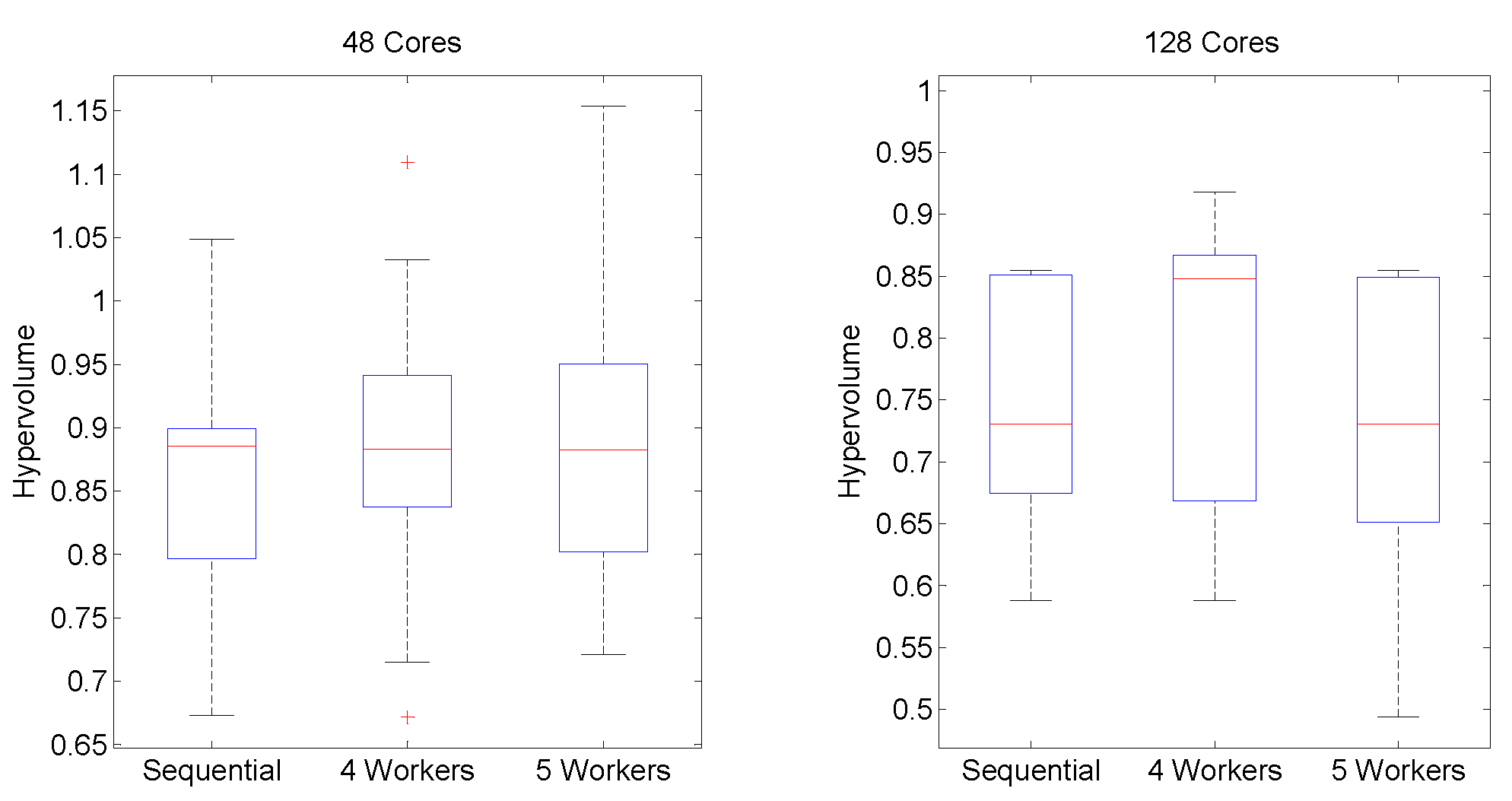}
\caption{Hypervolumes for 48 cores (left) and 128 cores(right) after 30 optimization runs. Both hypervolumes measured in the sequential and in the parallel execution, the latter with 4 and 5 workers.
The central line is the median, the edges of the box are the 25th and 75th percentiles, the whiskers extend to the most extreme data points not considered outliers, which are plotted individually (+ mark).}
\label{fig:hipervol48y128}
\end{figure}

The comparison has been carried out between the sequential execution, the 4-workers and the 5-workers versions of the parallel implementation. This choice was motivated because these configurations had obtained the highest non-saturated speed-ups. Results are shown in Figure \ref{fig:hipervol48y128}.
All were obtained after running 30 optimizations, each one with 250 generations in the 48 cores and 366 generation in the 128 cores.
As expected, the three boxplots inside each picture show a similar outcome; with 25th and 75th percentiles almost identical within the 48 and 128 core plots.

\subsection{Convergence of the MOEA}
\label{sec:convergenceMOEA}
\vspace{0.4cm}
The main benefit of parallelization is the considerable reduction in the fitness evaluation time and, consequently, in the the whole procedure of finding a good floorplan. 
In addition, we can take advantage of such a speed up for carrying out tests in order to detect possible weak points in the algorithm that, using the sequential version, would take months to complete.

As the population in an EA evolves, it is desirable to keep their diversity. 
Otherwise the exploration of the solution space will be guided towards a region, avoiding others which might be more promising. 
The analysis of the convergence is a straightforward method for verifying that diversity is maintained.
At the same time, it reveals whether the EA is well engineered or there is room for improvement. A slow convergence with good results is usually due to a poor representation or not appropriate genetic operators.

Although this work tackles a three objective problem, convergence is studied only for $J_2$ (wire length) and $J_3$ (thermal response) because $J_1>0$ means that none floorplan satisfies the constraints.

\begin{figure}[t]
\centering
\includegraphics[width=\textwidth, height=10cm]{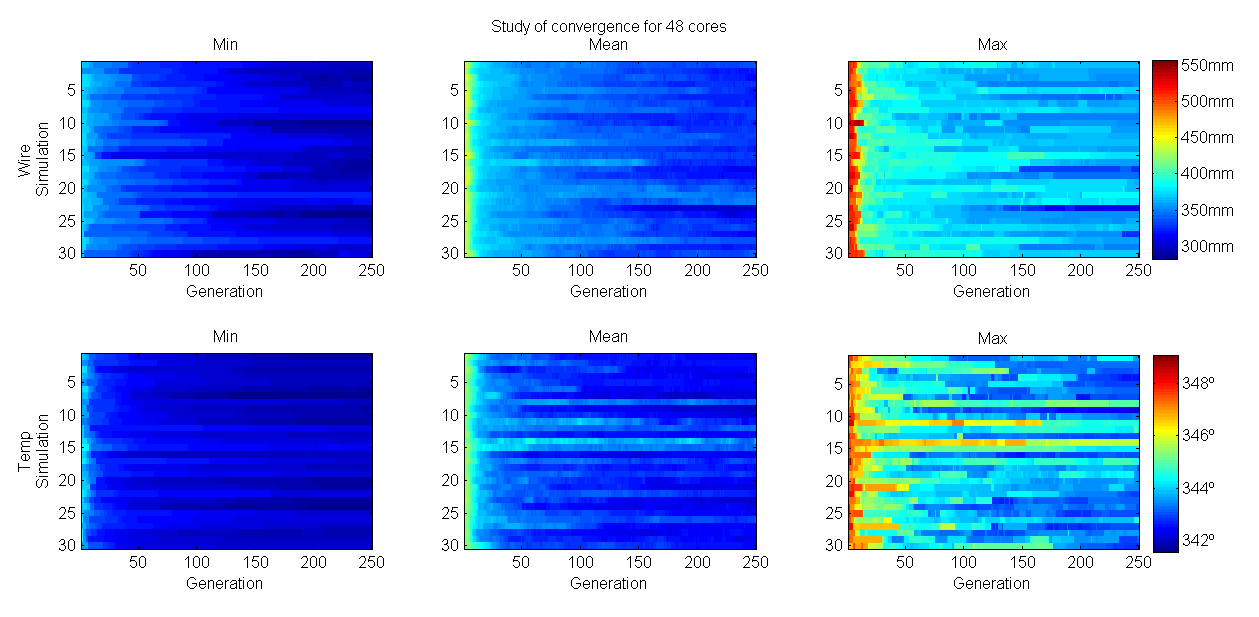}
\includegraphics[width=\textwidth, height=10cm]{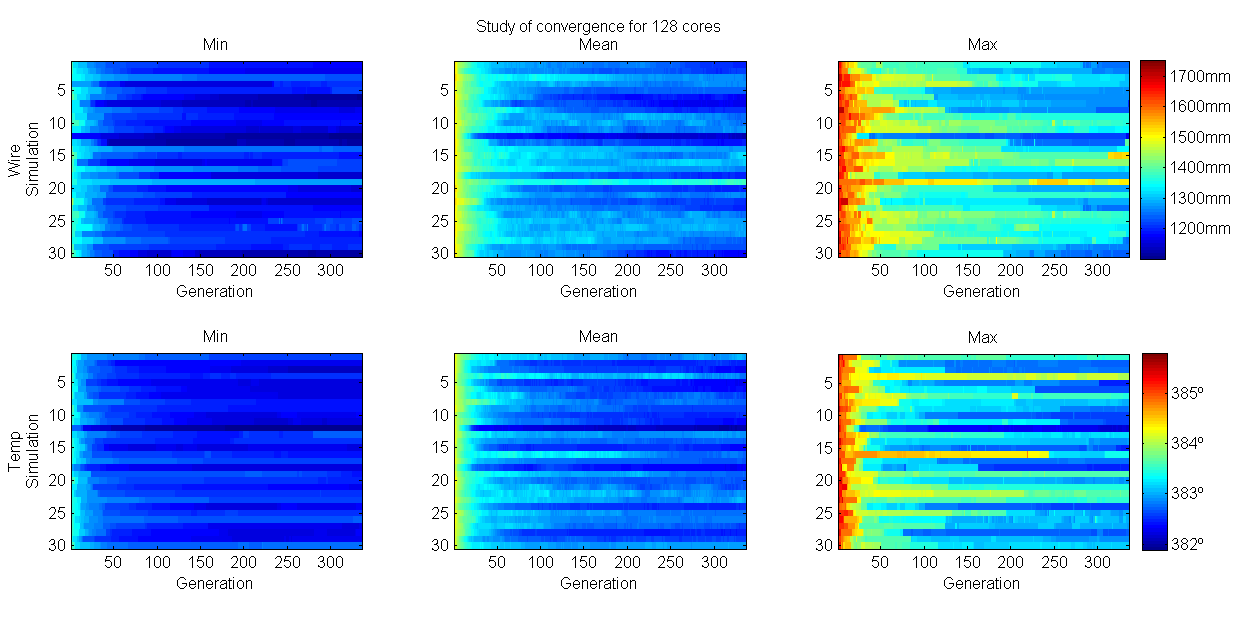}
\caption{Convergence evolution of the minimum, mean and maximum values for objectives $J_2$ (wire length) and $J_3$ (thermal response), considering only feasible individuals; for 48 cores (above) and 128 cores (below). }
\label{fig:converg48y128}
\end{figure}

Thus, values $J_2$ and $J_3$ of feasible solutions are extracted in arrays
$W_{r,g}$ and $T_{r,g}$ respectively, 
one for each optimization run $r=1\ldots30$ and each generation $g=1\ldots250$ for 48 cores and $g=1\ldots366$ for 128.
Then six matrices are constructed in the following way:
$$
\begin{array}{ccc}
{{\textbf{W}}_{{\textbf{min}}}}(r,g)  = \min  \left\{ {{W_{r,g}}} \right\}, &
{{\textbf{W}}_{{\textbf{mean}}}}(r,g) = \text{mean} \left\{ {{W_{r,g}}}, \right\} &
{{\textbf{W}}_{{\textbf{max}}}}(r,g)  = \max  \left\{ {{W_{r,g}}} \right\}, \\
{{\textbf{T}}_{{\textbf{min}}}}(r,g)  = \min  \left\{ {{T_{r,g}}} \right\}, &
{{\textbf{T}}_{{\textbf{mean}}}}(r,g) = \text{mean} \left\{ {{T_{r,g}}}, \right\} &
{{\textbf{T}}_{{\textbf{max}}}}(r,g)  = \max  \left\{ {{T_{r,g}}} \right\}. \\
\end{array}
$$
This procedure is done for both 48 and 128 core configurations. 
Finally, all six matrices are scaled between the minimum and maximum wire length and thermal response respectively, and plotted as shown in Figure \ref{fig:converg48y128}.

The left-most pictures, corresponding to the minimum values in each generation and optimization run of $J_2$ and $J_3$, show a decreasing behavior  eventually reaching the global minimum in almost all optimization runs for 48 cores and, at much slower rate, for 128 cores.
The middle pictures show convergence of the mean values of $J_2$ and $J_3$. Their trend is decreasing, starting at $1/2$ of the upper bound down to $1/4$ in the best cases and $1/3$ in the worst.
Finally, the convergence of maximum values of $J_2$ and $J_3$ is shown in the right-most pictures. Values decrease down to $1/3$ of the upper bound in most of the optimization runs.
To the light of these results it is clear that generations are better fitted as evolution advances. 
Moreover, since the mean of the last generation is quite close to the maximum in both objectives, less fitted individuals still have a considerable probability of being chosen, attesting that diversity is maintained. 
On the other hand, the slow convergence of the mean, especially in the larger problem of 128 cores, indicates that the representation could be better engineered. Genetic operators were discarded as reason because the offspring is always valid. Thus, no reconstruction is needed, which might lead to repeat certain schemas. 


\subsection{Thermal analysis}
\vspace{0.4cm}
Finally, we analyze the thermal optimization obtained with our algorithm. The floorplanner works with a fixed die size and aims to minimize both the total wire length and the maximum temperature of the chip. As we want to perform a thermal optimization of the described platforms, we need to provide the power consumption values and the areas of the different elements of the architectures as inputs to the thermal-aware floorplanner. In \cite{SPARC} we find that the power consumption of the {\sc sparc} is 4W at 1.4GHz. In the case of the {\sc Power6}, we find that 2.6W is the estimated power dissipation \cite{power6}. We consider the following areas: $3.24mm^2$ and $1.5mm^2$ for the {\sc sparc} and {\sc Power6} respectively (see \cite{SPARC} and \cite{power6}). The power consumption values and areas of the memories are found with the CACTI software \cite{cacti}. 

\begin{figure*}[!h]
	\includegraphics[height=2.1in,width=0.5\textwidth]{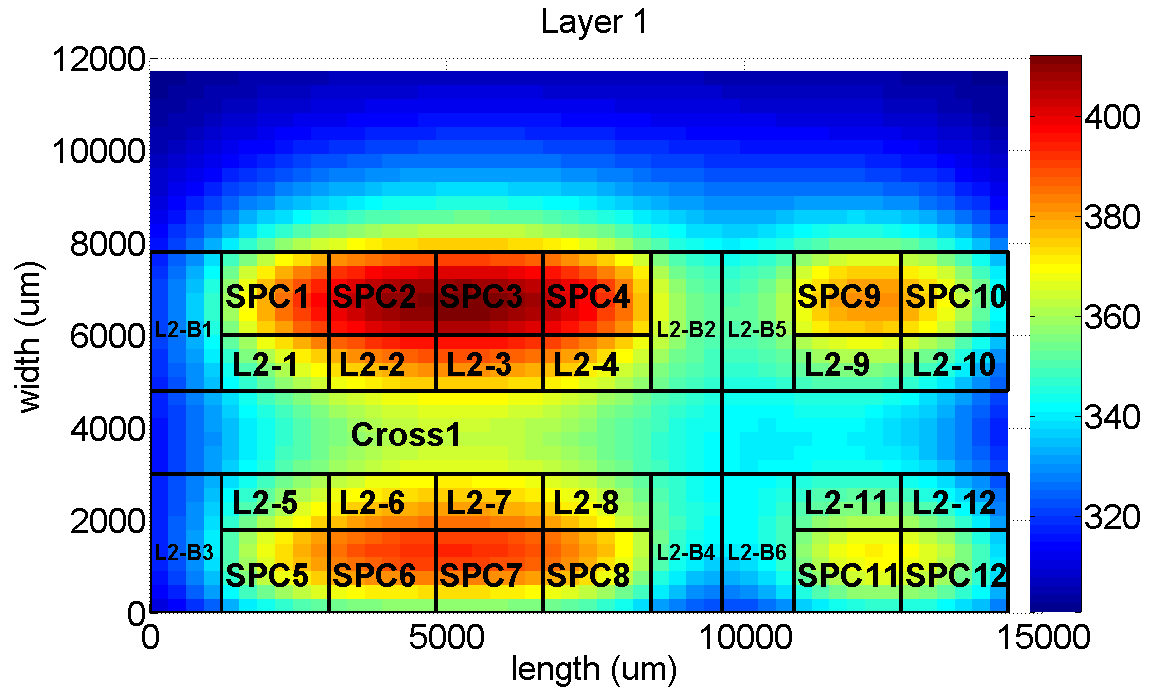}       						\includegraphics[height=2.1in,width=0.5\textwidth]{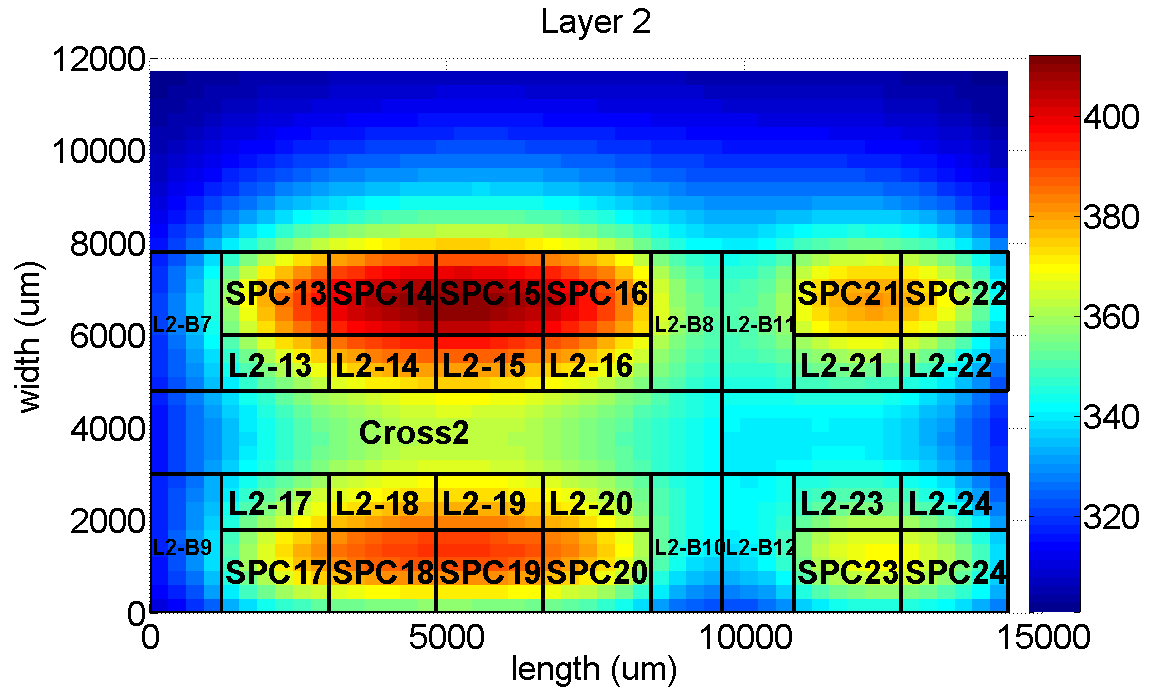} \\
    \includegraphics[height=2.1in,width=0.5\textwidth]{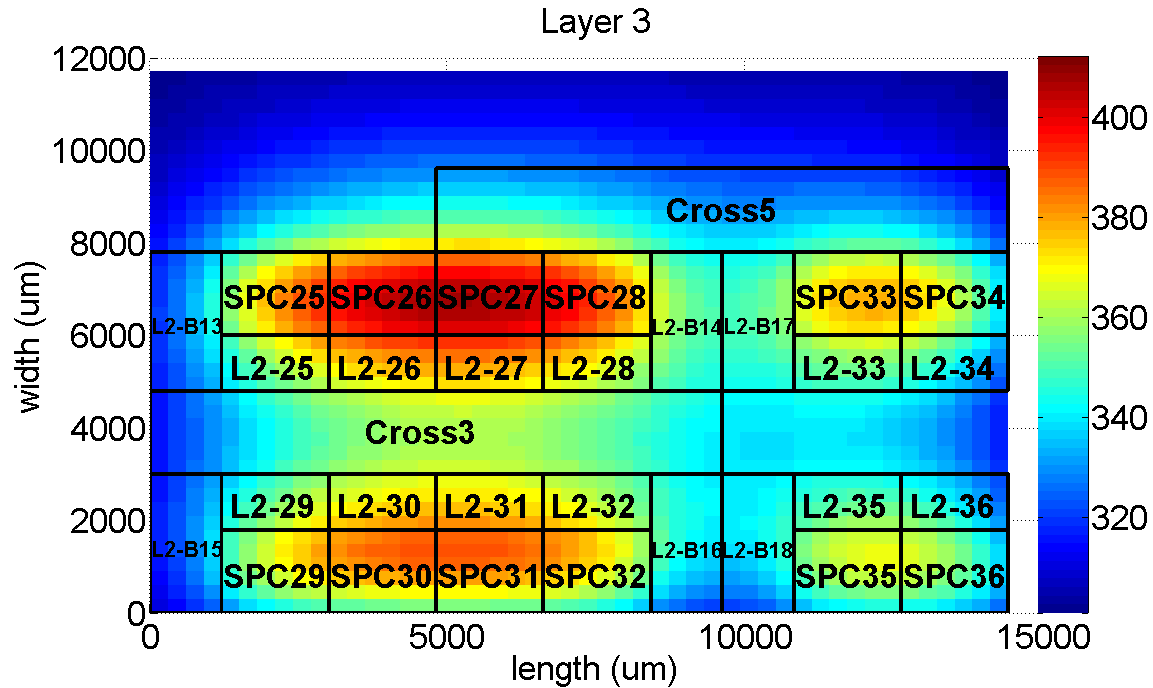}        						\includegraphics[height=2.1in,width=0.5\textwidth]{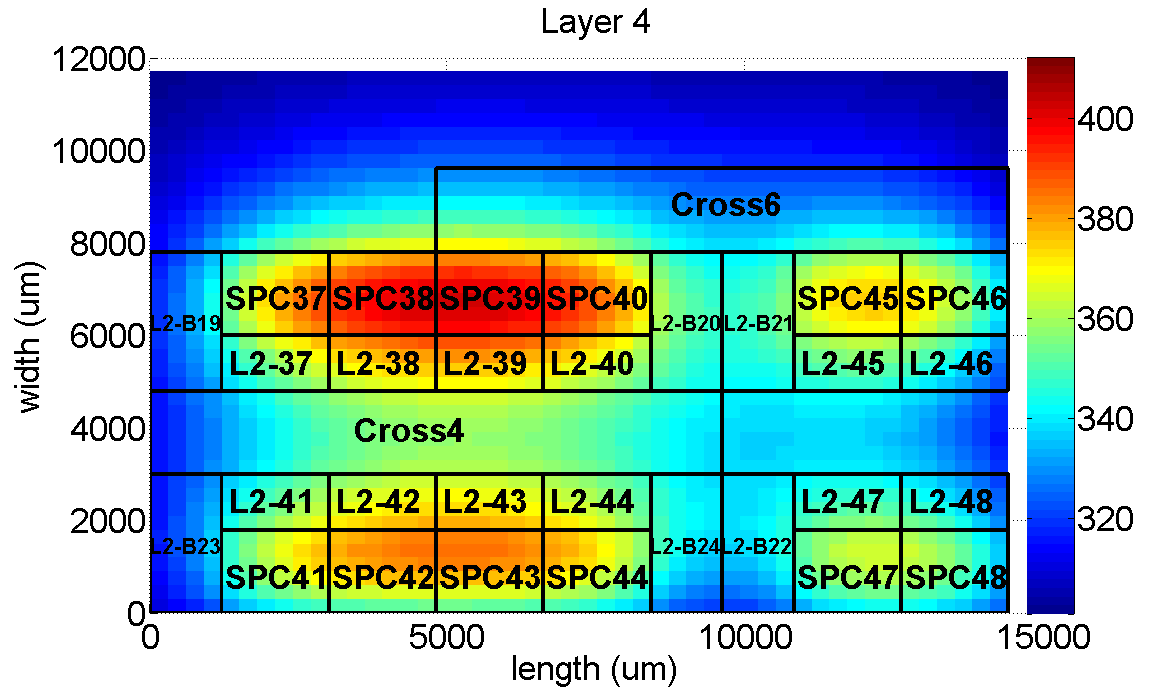}       
\caption{Thermal map of the 4 layers of the baseline configuration of the 48 cores platform.}
\label{thermalmaps48or}
\end{figure*}

\begin{figure*}[!h]
	\includegraphics[height=2.1in,width=0.5\textwidth]{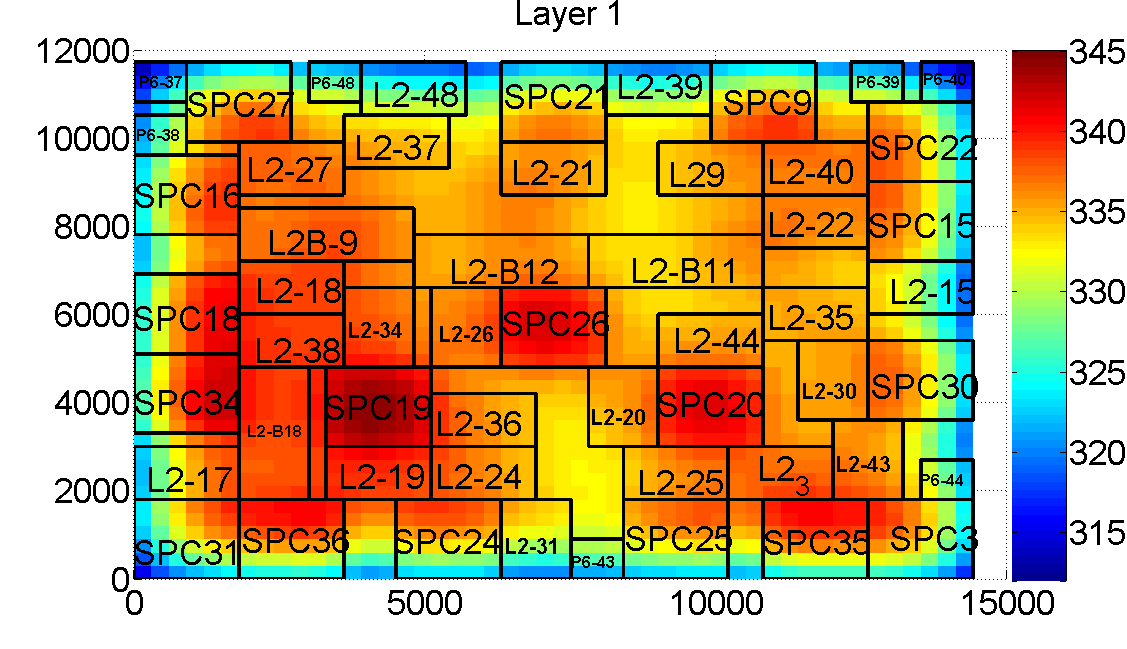}       							\includegraphics[height=2.1in,width=0.5\textwidth]{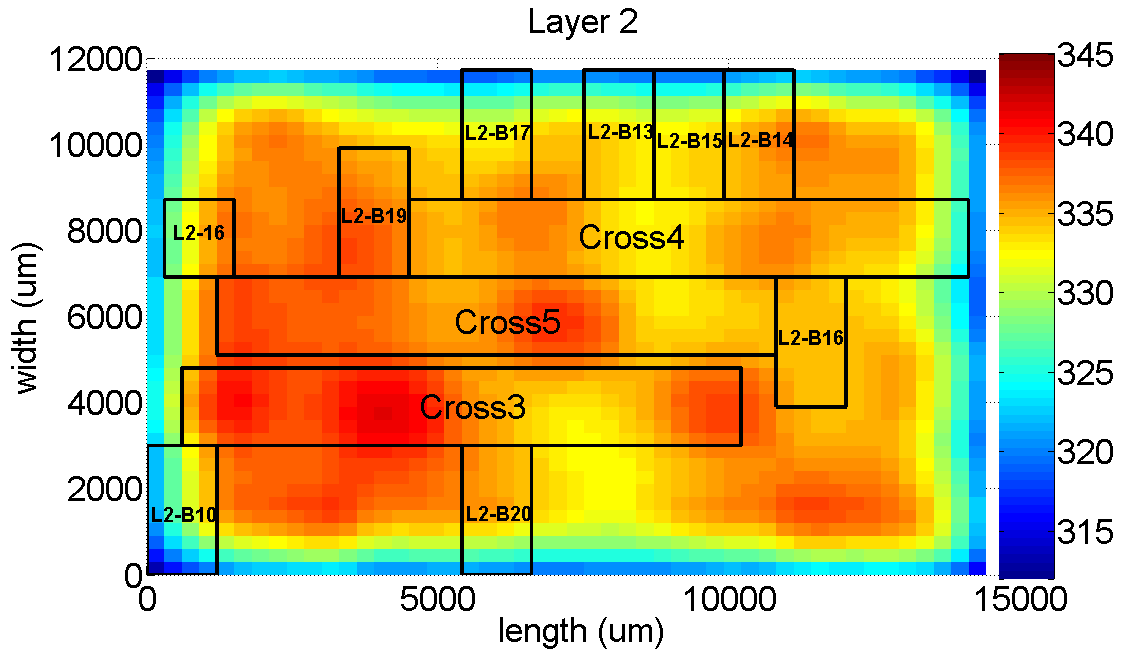} \\
	\includegraphics[height=2.1in,width=0.5\textwidth]{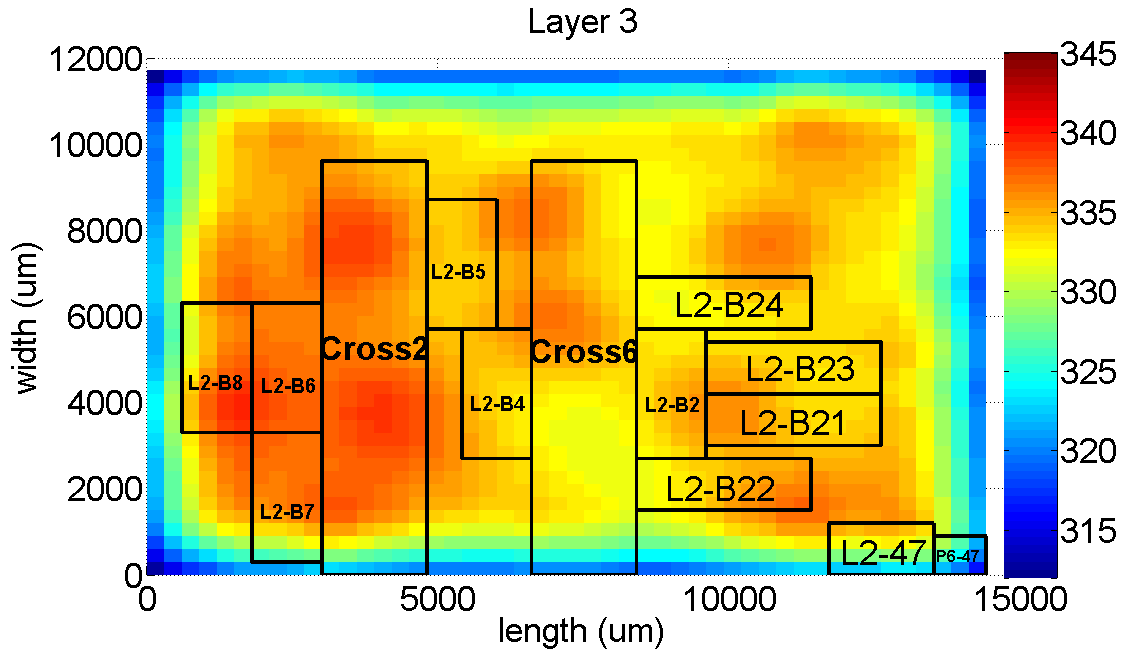}
	\includegraphics[height=2.1in,width=0.5\textwidth]{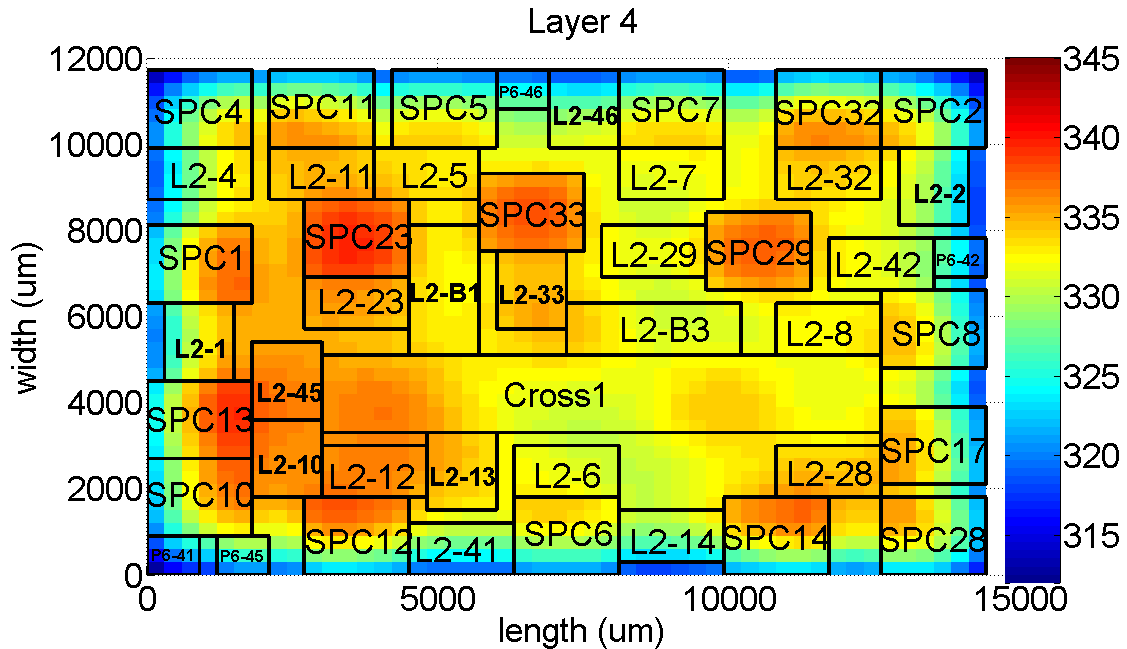}       
\caption{Thermal map of the 4 layers of a non-dominated solution of the 48 cores platform.}
\label{thermalmaps48op}
\end{figure*}

\vspace{0.4cm}
\textbf{48-cores configuration.~}
We compare an optimized configuration of the 48-cores heterogeneous platform to the 48-cores homogeneous platform represented in Figure \ref{thermalmaps48or}. In this baseline configuration, an original architecture composed of 12 cores is replicated in all the layers. As a consequence, the SPARC cores (SPC) are placed above the others producing hotspots. On the other hand, Figure \ref{thermalmaps48op} shows the thermal maps of the different layers of a non-dominated solution returned by the thermal-aware floorplanner. This figure shows an optimized placement of the SPARC cores (SPC), Power6 cores (P6), memories (L2) and crossbars (Cross) achieved by the floorplanner. In this configuration the hottest elements (SPARC cores) are generally placed in the borders of the chip and in the outer layers, separated as much as possible. In fact, the floorplanner avoids placing cores above the others as vertical heat spread is also taken into account. The crossbars are placed in intermediate layers to minimize the wire length.

The metrics considered for the thermal analysis of these two platforms are the maximum and mean temperature of the chip and the maximum thermal gradient. In Table \ref{thermalresponse48} we present the thermal response of these two different configurations. These results show that our floorplanner proposes thermally optimized configurations. The peak temperature of 411.82K found in the original configuration is reduced to 345.30K while the mean temperature is reduced in 12.54K. We can see that the maximum thermal gradient of the optimized configuration is reduced from 109.75K to 31.81K. Therefore, not only the temperature of the chip is reduced but it is also more evenly distributed. On the other hand, the wire length of the optimized configuration is a 2.11\% greater than the original which translates into a small performance penalty.

\begin{table}
\caption{Thermal response of the 48 cores configurations.}
\centering
\tabsize
\begin{tabular}{ccccc}
\toprule
  & Wire L.	 & $T_{MAX}$	& $T_{MEAN}$ & $Grad_{MAX}$  \\
\midrule
48 baseline   			& 6733 			& 411.82 K	& 344.29 K	 & 109.75\\
48 opt.  			& 6875 			& 345.30 K	& 331.75 K	 & 31.81\\
\bottomrule
\end{tabular}
\label{thermalresponse48}
\end{table}

\begin{figure*}[!h]
	\includegraphics[height=2.2in,width=0.33\textwidth]{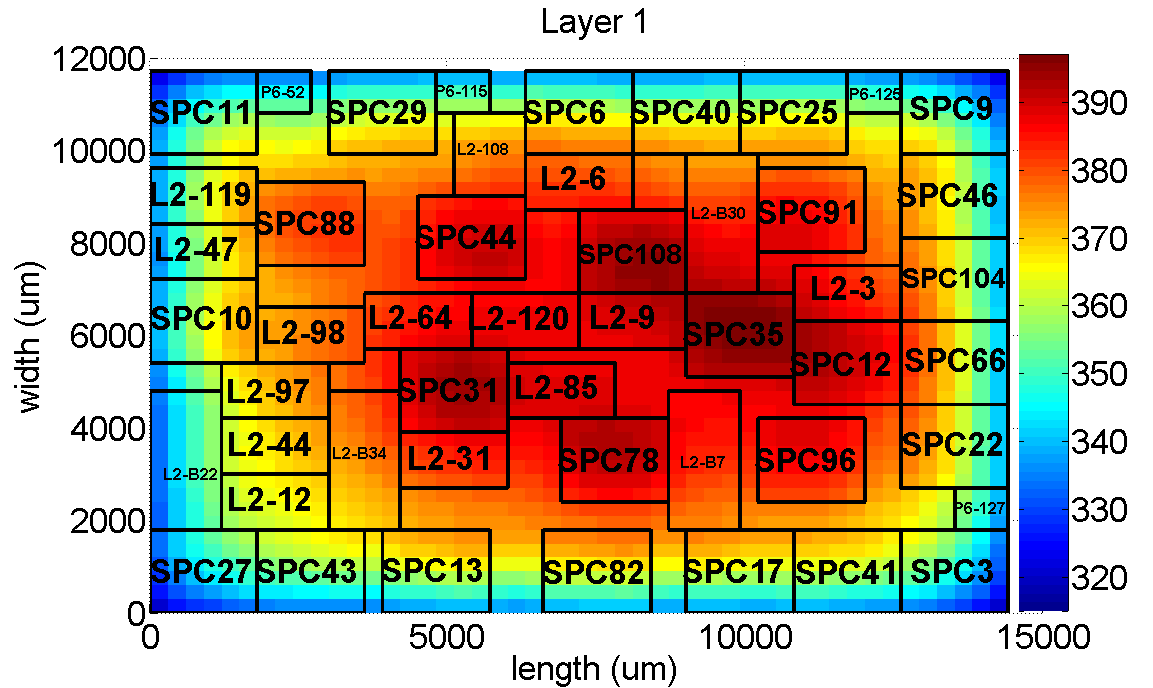}       						\includegraphics[height=2.2in,width=0.33\textwidth]{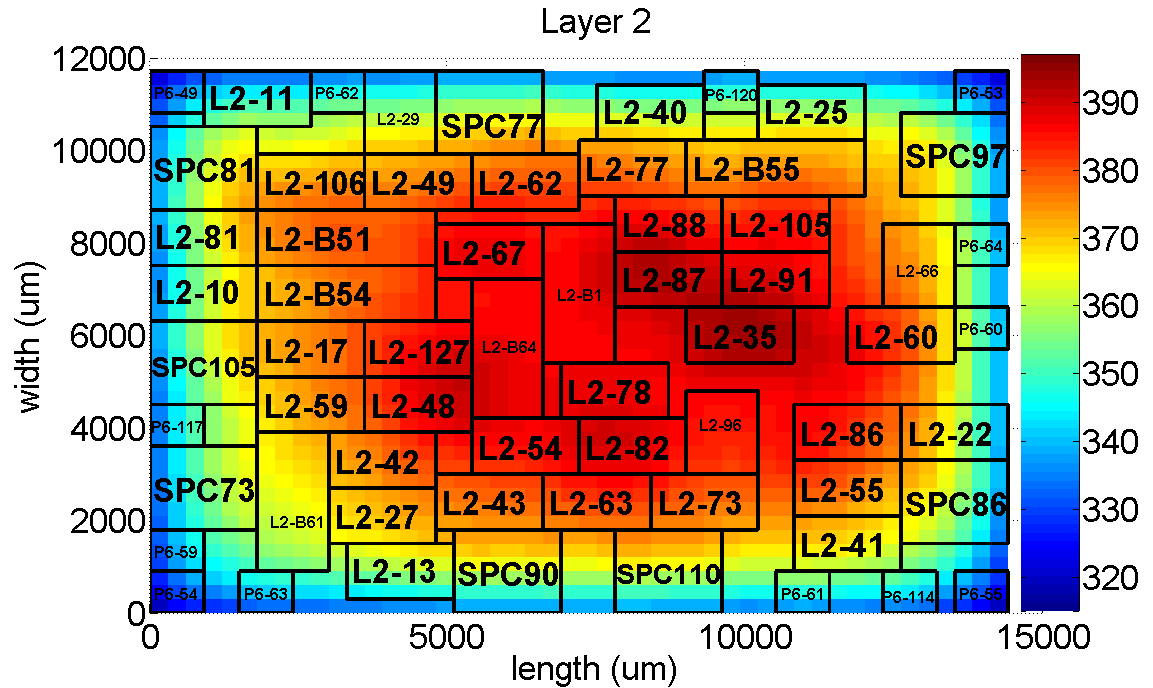} 								\includegraphics[height=2.2in,width=0.33\textwidth]{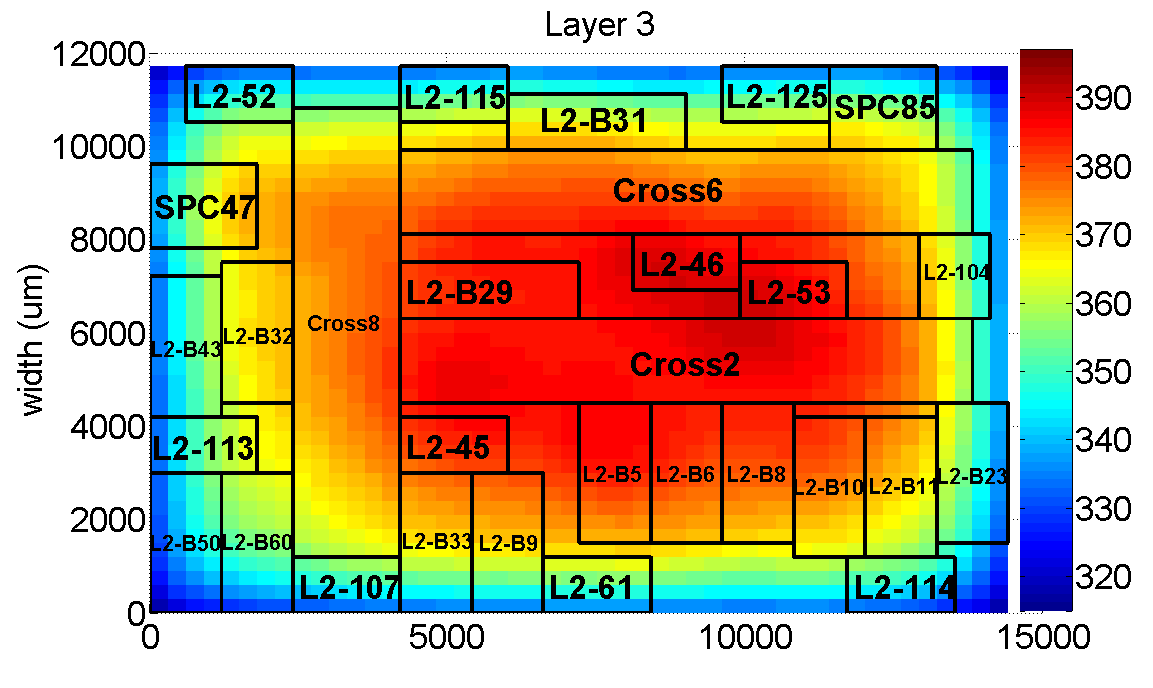} \\
    \includegraphics[height=2.2in,width=0.33\textwidth]{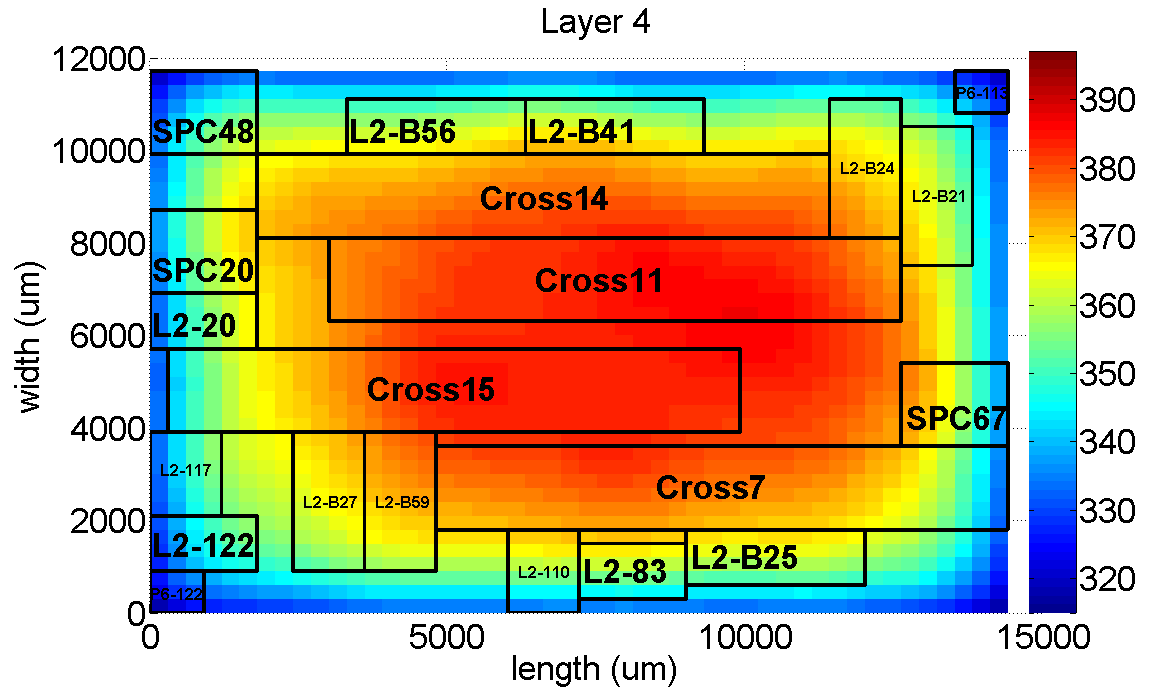} 								\includegraphics[height=2.2in,width=0.33\textwidth]{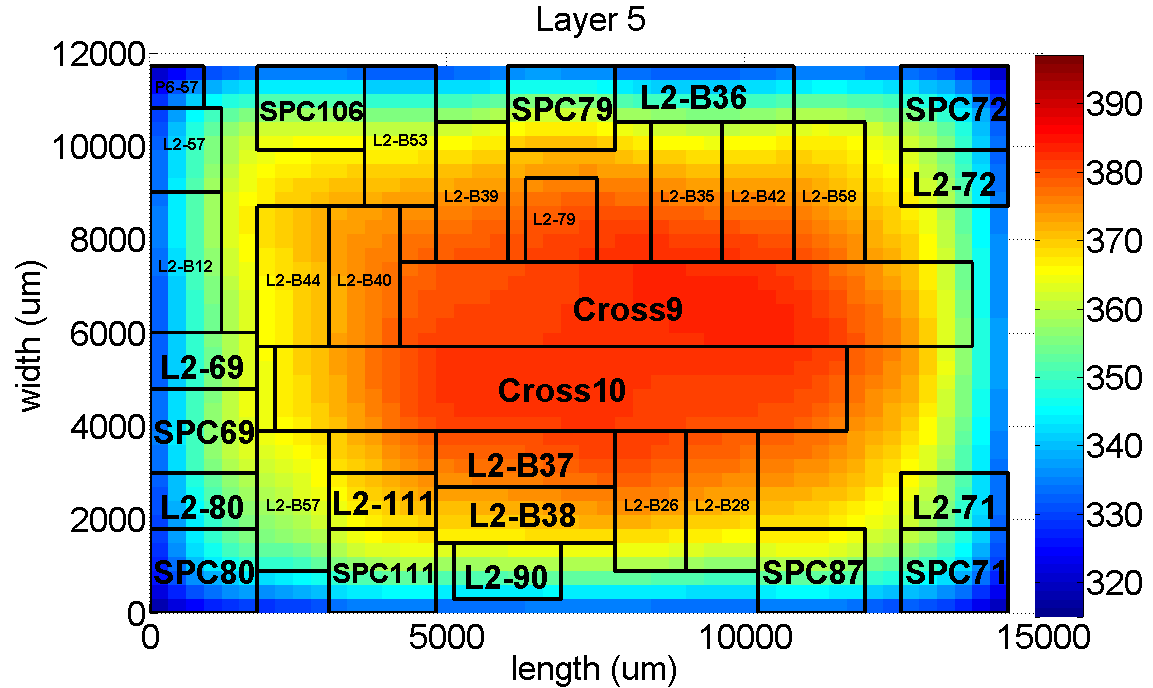}
    \includegraphics[height=2.2in,width=0.33\textwidth]{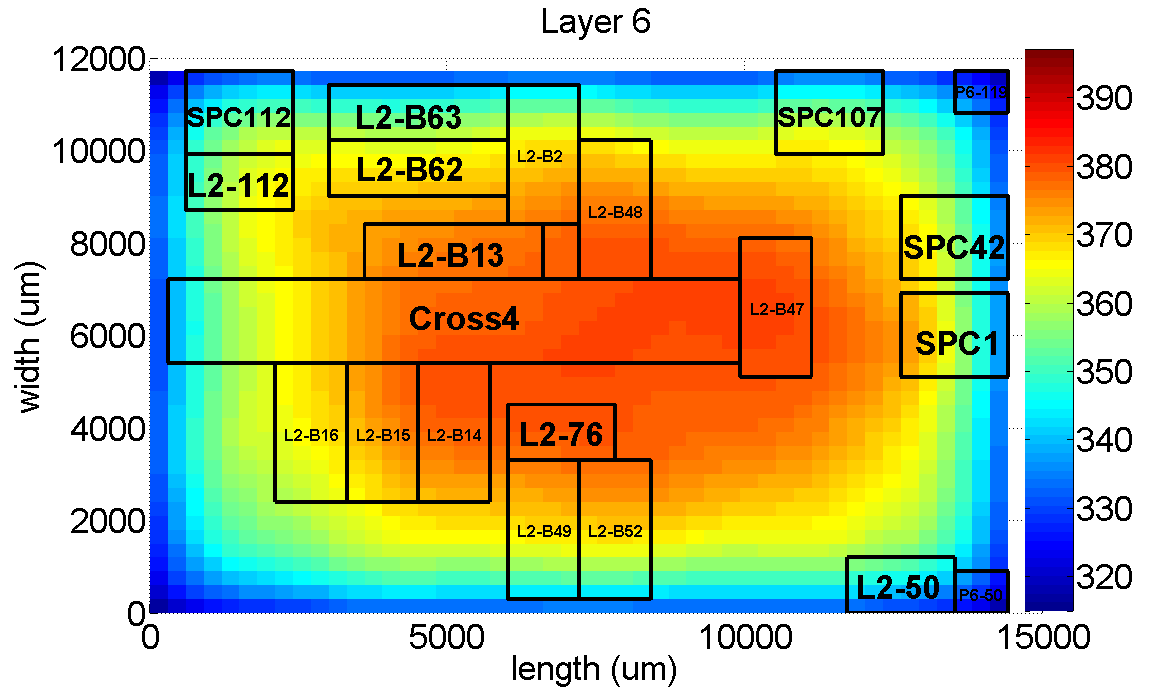} \\
    \includegraphics[height=2.2in,width=0.33\textwidth]{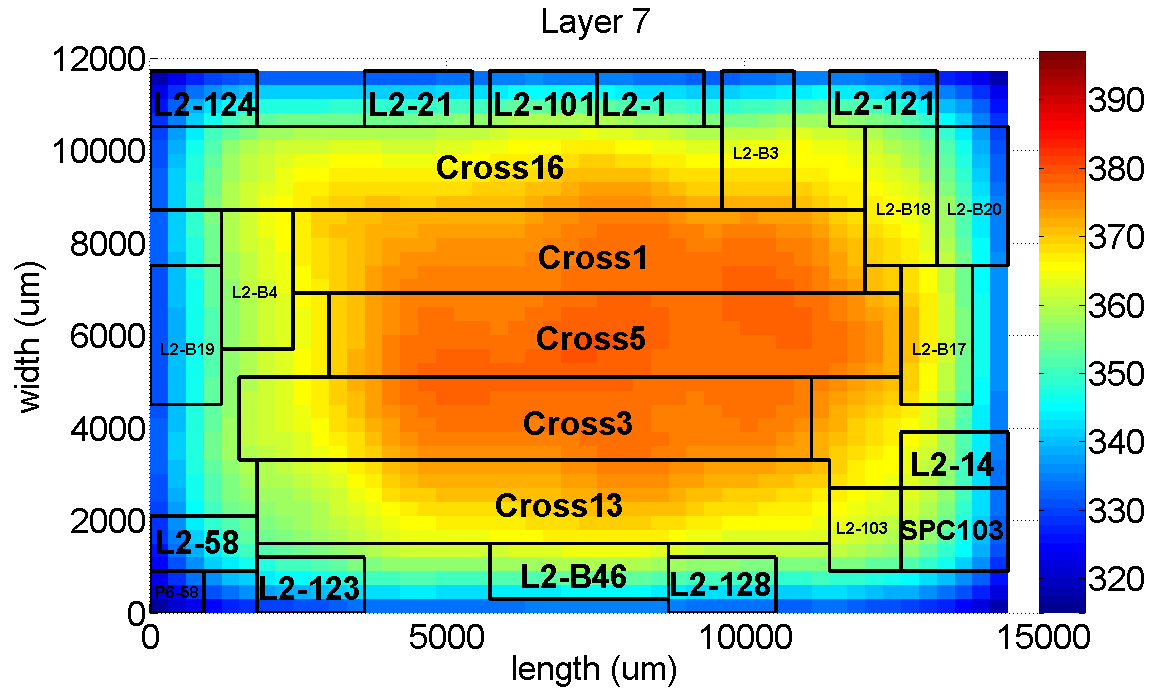}       
    \includegraphics[height=2.2in,width=0.33\textwidth]{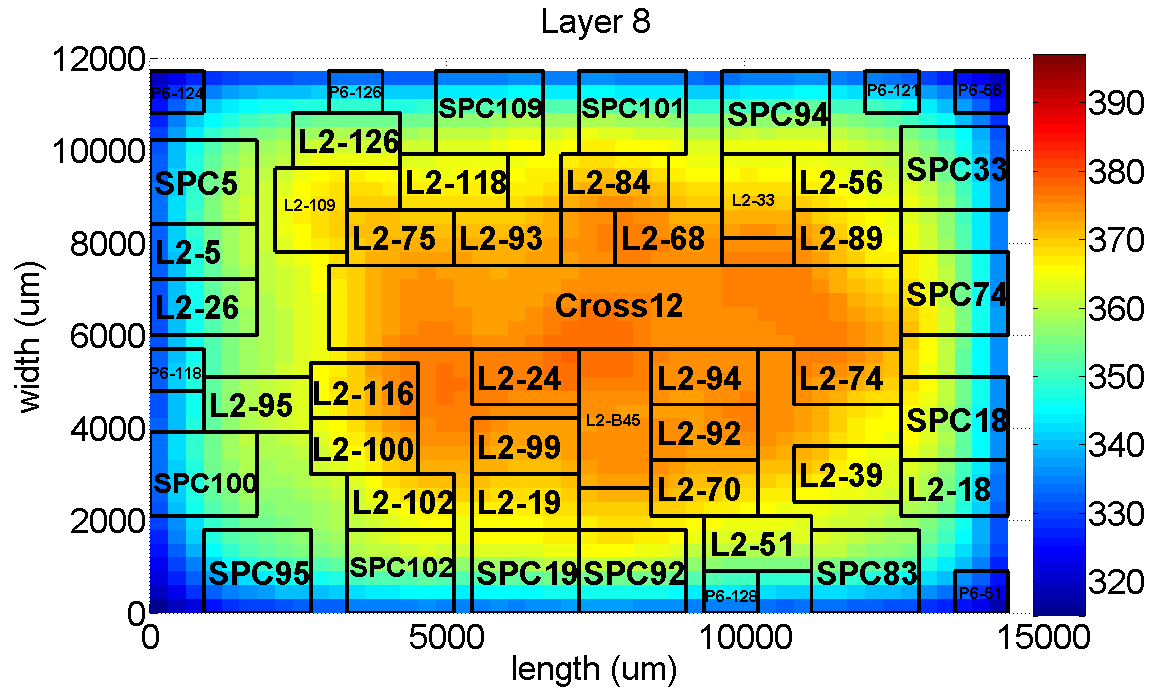}       
    \includegraphics[height=2.2in,width=0.33\textwidth]{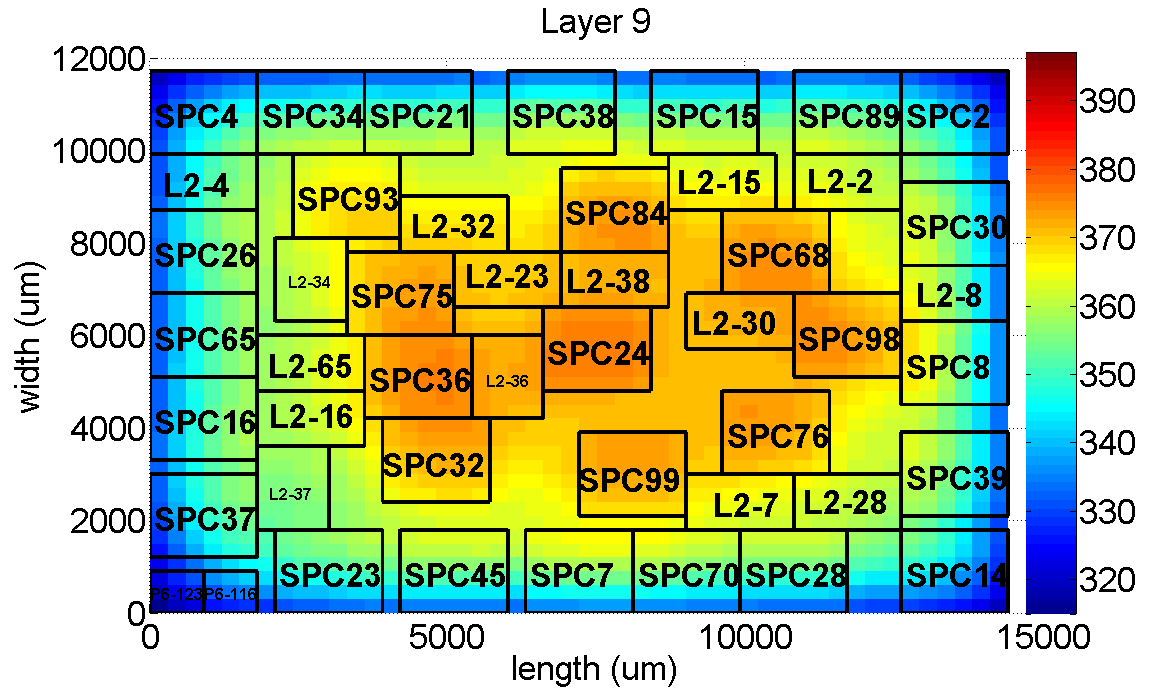} \\
    \caption{Thermal map of the 9 layers of a non-dominated solution of the 128 cores platform.}
\label{thermalmaps128}
\end{figure*}

\vspace{0.4cm}
\textbf{128-cores configuration.~}
For this larger configuration, we analyze one of the optimal floorplans obtained with our parallel implementation. Figure \ref{thermalmaps128} shows the thermal map of the chosen solution. As for the 48-cores platform, we can see that the SPARC cores tend to be placed in the outer layers and in the borders of the chip. The memories and the crossbars are placed in the inner layers. This way both the chip temperature and the wire length are minimized. Nevertheless hotspots appear in this configuration. Table \ref{thermalresponse128} shows the thermal response of an optimized configuration of the 128 cores platform. The hotspot visible in the first layer of the chip corresponds to the peak temperature of the chip reaching 396.84K. The mean temperature is 362.50K while the maximum thermal gradient is 75.80K. Further research and simulations with cooling techniques are required to study the feasibility of these architectures.

\begin{table}
\caption{Thermal response of the 128 cores configuration.}
\centering
\tabsize
\begin{tabular}{ccccc}
\toprule
	 & Wire L.		& $T_{MAX}$	& $T_{MEAN}$ & $Grad_{MAX}$ 	\\
\midrule
128 opt.  			 & 31587 		& 396.84 K	& 362.50 K	 & 75.80\\
\bottomrule
\end{tabular}
\label{thermalresponse128}
\end{table}


%% file: 04_conclusions_future.tex
\section{Conclusions and Future Work}
\vspace{0.4cm}

Current and short term future state-of-the-art in 3D many-core integration requires thermal-aware floorplanning techniques able to reduce peak and mean temperatures. However, current techniques that take into account thermal issues spend the most of the execution time dealing with decoding and evaluation of solutions.

This work has proposed a parallel implementation of a thermal-aware Multi-objective Evolutionary Algorithm for 3D floorplanning using a master-worker scheme. This model has provided optimized configurations for systems composed of 48 and 128 heterogeneous processor cores.

We have shown that the highest speedup values are obtained when the number of workers is closer to the number of cores of the processor that runs the algorithm. In our experiments, run on a 4-core processor, we have obtained maximum speedup values of 3.79 and 3.19 respectively for the 48 and 128 core test configurations by selecting 5 workers in the optimization algorithm. Furthermore, the parallelization presented in this work has made possible the study of the convergence of the floorplanner. The performed analysis shows that convergence evolves successfully in our experiments.

As a future work, we aim to overcome the drawbacks of the floorplanner presented in this work. First, we plan to replace the current approximated thermal model with a validated thermal simulator. In fact, the implemented thermal model is motivated by its low computational cost but it might not be obtaining accurately the thermal behavior of the different individuals. A recent work \cite{Arnaldo2012} has shown that the integration of a thermal simulator in the floorplanning process leads to a simultaneous reduction of the peak temperature and the wire length. Furthermore, the model proposed in the referred work is claimed to be, not only more accurate, but also faster than the approximated thermal model.

Another major improvement could be achieved with the use of a more suitable representation of the solutions. In fact, most of the floorplanning proposals are based on representations that require time consuming heuristics to decode the solutions. For instance, in our work, the decoding of the solutions together with the evaluation remains the bottleneck of the optimization process. A new representation allowing a direct mapping of the individuals into configurations of the architecture would alleviate the computational cost of the algorithm as the decoding step would be avoided. Furthermore, it would eliminate heuristics that might limit the exploration space and cause premature convergence problems. Thus, such a representation would be more suitable for fixed-outline floorplanning problems.

To propose a tool consistent with the state of the art of 3D chip design, thermal-aware floorplanners have to be implemented in accordance with current thermal simulators that split the IC into thermal cells (as done by 3D-ICE \cite{Sridhar2010}). This way, the coding of the solutions has to respect the grid-like structure used by this kind of simulators. Therefore, the thermal error due to the different cell sizes used in the optimization and validation processes is eliminated. Thus, a better thermal optimization can be achieved. The use of a grid-like representation together with the removal of placement heuristics makes the process well adapted for execution in massively parallel architectures such as GPUs.

A deeper study of new representations is our current and future work, and the preliminary results are very promising.

%% file: main_paper_CC_2012.bbl
\begin{thebibliography}{10}
\providecommand{\url}[1]{\texttt{#1}}
\providecommand{\urlprefix}{URL }
\expandafter\ifx\csname urlstyle\endcsname\relax
  \providecommand{\doi}[1]{doi:\discretionary{}{}{}#1}\else
  \providecommand{\doi}{doi:\discretionary{}{}{}\begingroup
  \urlstyle{rm}\Url}\fi

\bibitem{intelscc}
Intel. http://techresearch.intel.com/ProjectDetails.aspx?Id=1 2012.

\bibitem{Loh2010}
Loh GH, Xie Y. {3D} stacked microprocessor: Are we there yet? \emph{Micro,
  IEEE}  may-june 2010; \textbf{30}(3):60 --64, \doi{10.1109/MM.2010.45}.

\bibitem{Borkar1999}
Borkar S. Design challenges of technology scaling. \emph{Micro, IEEE}  1999;
  \textbf{19}(4), \doi{10.1109/40.782564}.

\bibitem{Srinivasan2004}
Srinivasan J, Adve S, Bose P, Rivers J. The impact of technology scaling on
  lifetime reliability. \emph{Dependable Systems and Networks, 2004
  International Conference on}, 2004; 177 -- 186,
  \doi{10.1109/DSN.2004.1311888}.

\bibitem{Sanka2005}
Sankaranarayanan K, \emph{et~al.}. A case for thermal-aware floorplanning at
  the microarchitectural level. \emph{Journal of Instruction-Level Parallelism}
   2005; \textbf{7}(1):8--16.

\bibitem{Berntsson2004}
Berntsson J, Tang M. A slicing structure representation for the multi-layer
  floorplan layout problem. \emph{EvoWorkshops}, 2004; 188--197.

\bibitem{Cong2004}
Cong J, Wei J, Zhang Y. A thermal-driven floorplanning algorithm for {3D ICs}.
  \emph{Proceedings of the 2004 IEEE/ACM International conference on
  Computer-aided design}, ICCAD '04, IEEE Computer Society: Washington, DC,
  USA, 2004; 306--313, \doi{http://dx.doi.org/10.1109/ICCAD.2004.1382591}.

\bibitem{Tang2007}
Tang M, \emph{et~al.}. A memetic algorithm for {VLSI} floorplanning. \emph{IEEE
  Transactions on Systems, Man, and Cybernetics, Part B}  2007;
  \textbf{37}(1):62--69.

\bibitem{Hung2005}
Hung WL, Xie Y, Vijaykrishnan N, Addo-Quaye C, Theocharides T, Irwin MJ.
  Thermal-aware floorplanning using genetic algorithms. \emph{Proceedings of
  the 6th International Symposium on Quality of Electronic Design}, ISQED '05,
  IEEE Computer Society: Washington, DC, USA, 2005; 634--639,
  \doi{http://dx.doi.org/10.1109/ISQED.2005.122}.

\bibitem{Han2007}
Han Y, Koren I. Simulated annealing based temperature aware floorplanning.
  \emph{J. Low Power Electronics}  2007; \textbf{3}(2):141--155,
  \doi{http://dx.doi.org/10.1166/jolpe.2007.128}.

\bibitem{Healy2007}
Healy M, \emph{et~al.}. Multiobjective microarchitectural floorplanning for
  2{D} and 3{D} {IC}s. \emph{CADICS, IEEE Transactions on}  2007;
  \textbf{26}(1):38--52, \doi{10.1109/TCAD.2006.883925}.

\bibitem{Cuesta2011b}
Cuesta~G\'{o}mez D, Risco~Mart\'{\i}n JL, Ayala JL, Hidalgo JI. A combination
  of evolutionary algorithm and mathematical programming for the {3D}
  thermal-aware floorplanning problem. \emph{Proceedings of the 13th annual
  conference on Genetic and evolutionary computation}, GECCO '11, ACM: New
  York, NY, USA, 2011; 1731--1738,
  \doi{http://doi.acm.org/10.1145/2001576.2001809}.

\bibitem{risco2011}
Arnaldo I, Risco-Mart\'{\i}n JL, Ayala JL, Hidalgo JI. Power profiling-guided
  floorplanner for thermal optimization in {3D} multiprocessor architectures.
  \emph{Proceedings of the 21st international conference on Integrated circuit
  and system design: power and timing modeling, optimization, and simulation},
  PATMOS'11, Springer-Verlag: Berlin, Heidelberg, 2011; 11--21.

\bibitem{Cantu98}
Cantú-Paz E. A survey of parallel genetic algorithms. \emph{Calculateurs
  Paralleles Reseaux et Systems Repartis}  1998; \textbf{10}:3289--3293.

\bibitem{VanVeldhuizen2003}
Van~Veldhuizen D, Zydallis J, Lamont G. Considerations in engineering parallel
  multiobjective evolutionary algorithms. \emph{Evolutionary Computation, IEEE
  Transactions on}  april 2003; \textbf{7}(2):144 -- 173,
  \doi{10.1109/TEVC.2003.810751}.

\bibitem{LopezJaimes2005}
Lopez~Jaimes A, Coello~Coello C. {MRMOGA}: parallel evolutionary multiobjective
  optimization using multiple resolutions. \emph{Evolutionary Computation,
  2005. The 2005 IEEE Congress on}, vol.~3, 2005; 2294 -- 2301 Vol. 3,
  \doi{10.1109/CEC.2005.1554980}.

\bibitem{Deb2002}
Deb K, \emph{et~al.}. A fast and elitist multiobjective genetic algorithm:
  {NSGA-II}. \emph{IEEE Transactions on Evolutionary Computation}  2002;
  \textbf{6}(2):182--197.

\bibitem{Paci2007}
Paci G, \emph{et~al.}. Exploring temperature-aware design in low-power mpsocs.
  \emph{International journal of embedded systems}  2007; \textbf{3}(1):43--51.

\bibitem{Zitzler1999}
Zitzler E, Thiele L. {Multiobjective Evolutionary Algorithms: A Comparative
  Case Study and the Strength Pareto Approach}. \emph{IEEE Transactions on
  Evolutionary Computation}  1999; \textbf{3}(4):257--271.

\bibitem{SPARC}
OpenSPARC. http://www.opensparc.net/pubs/preszo/07/n2isscc.pdf 2007.

\bibitem{power6}
IBM. http://www.ibm.com/systems/support/tools/estimator/energy 2012.

\bibitem{cacti}
HPlabs. www.hpl.hp.com/research/cacti/ 2012.

\bibitem{Arnaldo2012}
Arnaldo I, Vincenzi A, Ayala JL, Risco JL, Hidalgo JI, Ruggiero M, Atienza D.
  {Fast and Scalable Temperature-driven Floorplan Design in 3D MPSoCs}.
  \emph{13th Latin American Test Workshop (LATW)}, 2012.

\bibitem{Sridhar2010}
Sridhar A, Vincenzi A, Ruggiero M, Brunschwiler T, Atienza D. {3D-ICE}: Fast
  compact transient thermal modeling for {3D ICs} with inter-tier liquid
  cooling. \emph{Computer-Aided Design (ICCAD), 2010 IEEE/ACM International
  Conference on}, 2010; 463 --470, \doi{10.1109/ICCAD.2010.5653749}.

\end{thebibliography}
